\documentclass[9pt, journal, compsocconf]{IEEEtran}

\ifCLASSINFOpdf

\else

\fi

\usepackage{amsmath}
\usepackage{amsthm}
\usepackage{amssymb}

\usepackage{amsfonts}
\usepackage[ruled,lined]{algorithm2e}

\usepackage{graphicx}
\usepackage{subfigure}
\usepackage{tikz}
\usepackage{bm}
\usepackage{booktabs}
\usepackage{tabularx}
\usepackage{caption}
\usepackage{multirow}
\usepackage{rotating} %[figuresright]
\usepackage{lscape}

\usepackage{url}

\usepackage{tikz,mathpazo}
\usetikzlibrary{shapes.geometric, arrows}

\begin{document}

\title{Green Data Centers: A Survey, Perspectives, and Future Directions}

\author{\IEEEauthorblockN{Xibo Jin$^{1,2}$, Fa Zhang$^{1}$, Athanasios V. Vasilakos$^{3}$ and Zhiyong Liu$^{1}$}\\
\IEEEauthorblockA{
$^1$Institute of Computing Technology, Chinese Academy of Sciences\\
$^2$University of Chinese Academy of Sciences, Beijing, China\\
$^3$University of Western Macedonia, Greece\\
Email: $^{1,2}$\{jinxibo, zhangfa, zyliu\}@ict.ac.cn}  $^3$vasilako@ath.forthnet.gr
}

\maketitle

\begin{abstract}
At present, a major concern regarding data centers is their extremely high energy consumption and carbon dioxide emissions. However, because of the over-provisioning of resources, the utilization of existing data centers is, in fact, remarkably low, leading to considerable energy waste. Therefore, over the past few years, many research efforts have been devoted to increasing efficiency for the construction of green data centers. The goal of these efforts is to efficiently utilize available resources and to reduce energy consumption and thermal cooling costs. In this paper, we provide a survey of the state-of-the-art research on green data center techniques, including energy efficiency, resource management, thermal control and green metrics. Additionally, we present a detailed comparison of the reviewed proposals. We further discuss the key challenges for future research and highlight some future research issues for addressing the problem of building green data centers.
\end{abstract}

\begin{IEEEkeywords}
Green data center; Energy efficiency; Resource management; Thermal control; Green metrics, monitoring and experimental techniques
\end{IEEEkeywords}

\IEEEpeerreviewmaketitle

\section{Introduction}
As hosts of large-scale service applications, data centers now play a crucial role in modern Information Technology (IT) infrastructures. With the development of cloud computing, data centers are growing exponentially in number and size. A new study from International Data Corporation (IDC) predicts that the total number of data centers deployed worldwide will peak at $8.6$ million in 2017 \cite{InternationalDataCorporation}. Consequently, the extremely high energy consumption and carbon dioxide emissions of data centers are becoming a major concern worldwide. The Natural Resources Defense Council (NRDC) reported that 91 billion KWh of electricity was consumed by U.S. data centers in 2013. Continuing this trend, NRDC estimates that the electricity usage will reach 140 billion KWh by 2020. As a result, data center electricity consumption will cost American businesses \$13 billion annually in electricity bills and result in the emission of nearly 100 million metric tons of carbon pollution per year \cite{NRDC}. Globally, it had revealed that power requirements grew by 63\% to 38 Gigawatts in 2012, up from 24 Gigawatts in 2011 \cite{VenkatramanGlobalDatacentercensus2012}. Moreover, data centers will consume 8\% of the worldwide electricity supply by 2020 while this fraction was about 1.3\% in 2010 \cite{GaoSIGCOMM12}. Therefore, making data centers ``green'' can reduce both costs for energy consumption and carbon dioxide emissions.

Existing data centers often operate at low utilization because of over-provisioning and fragmentation of resources \cite{Greenberg09, Patel08}, leading to considerable energy waste. According to a McKinsey study in 2008 \cite{Kaplan08}, the typical utilization ratio is approximately 6\%. A Gartner report from 2012 \cite{James12} found the typical utilization rates to be in the 7\% to 12\% range, slightly better than the result of the 2008 McKinsey study. Recently, Google Inc. reported that they can improve the utilization of their servers to relatively high rates of 20-40\% \cite{Barroso13}. Many studies \cite{Carrega12, Abts10, Benson10} have also shown that data center networks experience high underutilization, with typical utilization ranging between 5\% and 25\%. What makes the matter worse is that the low utilization of these servers and network resources also cause the waste of other supporting infrastructure, such as power distribution and cooling. From these data, we can see that the present energy efficiency of data centers is extremely low and that there is great potential for reducing energy consumption in data centers.

Motivated by the high energy consumption and low utilization of data centers, many research efforts over the past few years have focused on the design of green data center infrastructures and services. Generally, these approaches can be categorized into two major classes \cite{Rivoire09}: 1) those that adopt ``green'' equipment in the preliminary design and building phase of the data centers, and 2) those that increase the efficiency of the daily expenditures incurred during the operation of the data centers. In this paper, we focus on the latter class of methods, i.e., those that emphasize managing the procedures for running and operating data centers in a ``green'' manner. To this end, research studies on the development of green data centers have concentrated on the following aspects: 1) decreasing the power consumption of data center resources, 2) increasing the utilization of data centers, 3) controlling the thermal behavior of data centers, and 4) developing green metrics, monitoring and experimental techniques. From the perspective of data center architecture, the fundamental components of data centers are 1) computing servers, 2) connection networks, and 3) cooling equipment. Correspondingly, green data center technologies can be applied to a individual component or to a hybrid scheme of components.

\begin{figure*}[htbp]
\centering
\tikzset{
box/.style={rectangle, draw=black, align=center, inner sep=2pt}
}
\begin{tikzpicture}
  
  \node[box, line width=0.9pt] (n1) at (5.9,6.3) [text width=4cm, minimum height=0.9cm]{Green Data Center};
  
  \draw[line width=0.9pt] (0,5.2)--(12,5.2);
  
  \draw[line width=0.9pt] (6,5.85)--(6,5.2);
  
  \draw[line width=0.9pt] (0,5.2)--(0,4.7);
  \draw[line width=0.9pt] (4,5.2)--(4,4.7);
  \draw[line width=0.9pt] (8,5.2)--(8,4.7);
  \draw[line width=0.9pt] (12,5.2)--(12,4.7);
  
  \node[box, line width=0.9pt] (n2) at (0,4.2) [text width=2cm, minimum height=1cm]{Energy Efficiency};
  \node[box, line width=0.9pt] (n3) at (4,4.2) [text width=2.5cm, minimum height=1cm]{Resource Management};
  \node[box, line width=0.9pt] (n4) at (8,4.2) [text width=1.5cm, minimum height=1cm]{Thermal Control};
  \node[box, line width=0.9pt] (n5) at (12,4.2) [text width=1.5cm, minimum height=1cm]{Green Metrics};
  
  \draw[line width=0.9pt] (0,3.7)--(0,0.5);

  \draw[line width=0.9pt] (0,2.7)--(0.2,2.7);
  \draw[line width=0.9pt] (0,1.6)--(0.2,1.6);
  \draw[line width=0.9pt] (0,0.5)--(0.2,0.5);
  
  \node[box, line width=0.9pt] (n6) at (1.8,2.7) [text width=3cm, minimum height=0.9cm]{Dynamic Speed Scaling};
  \node[box, line width=0.9pt] (n7) at (1.8,1.6) [text width=3cm, minimum height=0.9cm]{Power-Down Mechanism};
  \node[box, line width=0.9pt] (n8) at (1.8,0.5) [text width=3cm, minimum height=0.9cm]{Hybrid Technology};
  
  \draw[line width=0.9pt] (4,3.7)--(4,0.0);

  \draw[line width=0.9pt] (4,2.7)--(4.2,2.7);
  \draw[line width=0.9pt] (4,1.8)--(4.2,1.8);
  \draw[line width=0.9pt] (4,0.9)--(4.2,0.9);
  \draw[line width=0.9pt] (4,0.0)--(4.2,0.0);

  \node[box, line width=0.9pt] (n9) at (5.8,3) [text width=3cm, minimum height=0.9cm]{Virtual Machine Assignment};
  \node[box, line width=0.9pt] (n10) at (5.8,2) [text width=3cm, minimum height=0.9cm]{Network Traffic Engineering};
  \node[box, line width=0.9pt] (n11) at (5.8,1) [text width=3cm, minimum height=0.9cm]{Power Distribution};
  \node[box, line width=0.9pt] (n12) at (5.8,0.0) [text width=3cm, minimum height=0.9cm]{Renewable Energy Access};
    
  \draw[line width=0.9pt] (8,3.7)--(8,1);

  \draw[line width=0.9pt] (8,2.5)--(8.2,2.5);
  \draw[line width=0.9pt] (8,1)--(8.2,1);
  
  \node[box, line width=0.9pt] (n13) at (9.8,2.5) [text width=3cm, minimum height=0.9cm]{Cooling and Workload Distribution};
  \node[box, line width=0.9pt] (n14) at (9.8,1) [text width=3cm, minimum height=0.9cm]{Temperature-Reliability Trade-off};
  
  \draw[line width=0.9pt] (12,3.7)--(12,1);

  \draw[line width=0.9pt] (12,2.5)--(12.2,2.5);
  \draw[line width=0.9pt] (12,1)--(12.2,1);
  
  \node[box, line width=0.9pt] (n15) at (13.8,2.5) [text width=3cm, minimum height=0.9cm]{Green Metrics};
  \node[box, line width=0.9pt] (n16) at (13.8,1) [text width=3cm, minimum height=0.9cm]{Green Monitoring and Experimental Techniques};
 
\end{tikzpicture}
\caption{\label{fig:An_overview_of_green_data_center_techniques}An overview of green data center techniques}
\end{figure*}
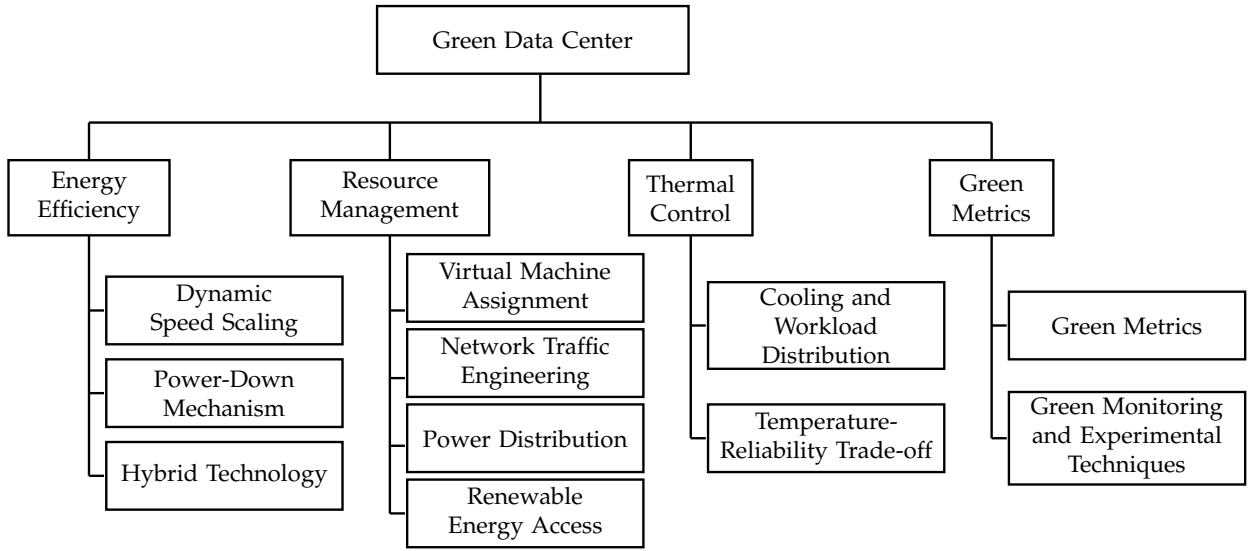

Although obvious progress has been made in the construction of green data centers over the past decade, there is still a large gap, and therefore a large opportunity for savings, between the current average and the characteristics of a best-practice green data center \cite{NRDC14}. In this paper, we present a survey of the current state-of-the-art research on green data centers, as summarized in Fig.~\ref{fig:An_overview_of_green_data_center_techniques}. Some recent surveys have conducted on energy-aware resource allocation for cloud computing \cite{BeloglazovFGCS2012} or green data center networks \cite{BilalFGCS2014}. Different from them, our survey covers many aspects of the development of green data centers. Our main contributions are two-fold: 1) we discuss the key insights underlying recent strategies and compare the relevant proposals, and 2) we note the future research challenges and directions for various aspects of the design of green data centers.

The remainder of this paper is organized as follows. First, we discuss the studies on and challenges of decreasing power consumption in Section~\ref{sec:Energy efficiency}. We introduce efforts toward and opportunities for the efficient utilization of data center resources in Section~\ref{sec:Data center resource management}. We explore studies and potential cooling issues related to data center temperature and thermal control in Section~\ref{sec:Temperature and thermal control}, and we describe the development of and guidelines for green metrics, monitoring and experimental techniques in Section~\ref{sec:Green metrics, monitoring and experimental techniques}. Finally, we conclude the paper in Section ~\ref{sec:Conclusion}.
\section{Energy Efficiency}\label{sec:Energy efficiency}
The direct approach to the development of green data centers is to curb their energy use by employing energy efficiency techniques. The fundamental concept is to exploit power management technology in the \textbf{I}nformation and \textbf{C}ommunication \textbf{T}echnology (ICT) devices of data centers. In this regard, the research community has proposed a number of possible approaches in recent years.
\subsection{Dynamic Speed Scaling (SS)}
This scheme, which is also known as Dynamic Voltage and Frequency Scaling (DVFS), focuses on lowering the frequency/speed of devices to save energy. Dynamic speed scaling allows for power savings as power consumption is approximately proportional to the supply voltage or the device speed $s$ (e.g., following the cube-root rule, $s^3$, or a more general function $f(s,\alpha)$, where $\alpha$ is a constant power parameter) \cite{Yao95, Mudge01, AlbersCACM10}. The goal is to determine the processing speeds and job assignment to minimize the total energy cost and guarantee the prescribed performance constraints. Intensive research, initiated by Yao \emph{et al.} \cite{Yao95}, has been conducted in pursuit of saving energy by speed scaling. Some efforts address the problem of scheduling via job deadlines \cite{LiPNAS06} and the optimization of job flow times \cite{AlbersATLG07}. Other approaches consider the problem of makespan minimization \cite{BundeScheduling09}, incorporate precedence constraints defined between jobs \cite{PruhsTCS08} and account for affinities between jobs and processors \cite{JinEuroPar13}. The research on variable-speed architectures includes single processors \cite{Yao95, BansalJACMM07}, parallel processors \cite{AlbersSPAA07, GreinerSPAA09}, and network devices \cite{GunaratneTOC08, AndrewsTON12, BampisFSTTCS13} with preemption or nonpreemption settings \cite{AntoniadisScheduling13, BampisDAM15}. Some studies also have addressed realistic speed models in which only a finite set of discrete speed levels is available \cite{LiJSIAM05, JinIPDPSForum13} or in which the speeds are bounded on an interval \cite{BansalICALP08}. In addition, Bampis \emph{et al.} \cite{BampisFSTTCS13} studied a heterogeneous multiprocessor preemptive problem, in which it was assumed that each processor had a different speed-to-power function. We can present all these efforts in a consistent fashion by extending the standard three-field notation of \cite{GrahamAnnalsDisMath79}; we summarize the algorithmic results obtained for dynamic speed scaling in Tables~\ref{tb:Algorithmic results of minimizing energy subject to deadline feasibility constraints on speed scaling} and \ref{tb:Algorithmic results of flow time, makespan, throughput on speed scaling}, and the symbols used are defined in Table~\ref{tb:Definitions of symbol on dynamic speed scaling}.

\begin{sidewaystable*}[htbp]
\footnotesize
\renewcommand{\arraystretch}{1.3}
\centering
\caption{Algorithmic results for speed scaling approaches based on the minimization of energy subject to deadline feasibility constraints.}
\begin{tabular}{c|c|c|c|c|c}
\hline\hline
\cline{1-6}
Speed Mode & Environment &   Problem   & Algorithm Type &   Complexity   &   Approximation/Competitive Ratio \\\hline
\multicolumn{1}{c|}{\multirow{29}{2.2cm}{Continuous speed scaling}}   &   \multicolumn{1}{c|}{\multirow{11}{2.2cm}{Single processor}}   &   $1{\vert}r_j$, $d_j$, $pmtn{\vert}E$   & \multicolumn{1}{c|}{\multirow{1}{2.3cm}{Offline}}  &$O(n^3)$ \cite{Yao95}, $O(n^2\log{n})$ \cite{LiPNAS06}& \verb'--'  \\
\cline{4-6}
\multicolumn{1}{c|}{}   &   \multicolumn{1}{c|}{}   &   & \multicolumn{1}{c|}{\multirow{2}{2.5cm}{Online}} & & $\alpha^{\alpha}$ \cite{Yao95}, $2(\frac{\alpha}{\alpha-1})^{\alpha}e^{\alpha}$ \cite{BansalJACMM07}   \\
\multicolumn{1}{c|}{}   &   \multicolumn{1}{c|}{}   &   &  \multicolumn{1}{c|}{}   & & ${4^{\alpha}}/({2{e^{\frac{1}{2}}\alpha^{\frac{1}{2}}}})$ \cite{BansalICALP09}   \\
\cline{4-6}
\multicolumn{1}{c|}{}   &   \multicolumn{1}{c|}{}   &   $1{\vert}agreeable$, $pmtn{\vert}E$   &  \multicolumn{1}{c|}{\multirow{1}{2.3cm}{Offline}}  &$O(n^2)$ \cite{WuTCS11}& \verb'--'  \\
\cline{3-6}
\multicolumn{1}{c|}{}   &   \multicolumn{1}{c|}{}   &   $1{\vert}r_j$, $d_j$, $w_j=1$, $non{\verb'-'}pmtn{\vert}E$   &   \multicolumn{1}{c|}{\multirow{7}{2.3cm}{Offline}}   &Polynomial \cite{HuangMFCS14}&  \verb'--'  \\
\multicolumn{1}{c|}{}   &   \multicolumn{1}{c|}{}   &   $1{\vert}agreeable$, $non{\verb'-'}pmtn{\vert}E$   & \multicolumn{1}{c|}{}  &   Polynomial \cite{Yao95, AntoniadisScheduling13}   &  \verb'--'  \\
\multicolumn{1}{c|}{}   &   \multicolumn{1}{c|}{}   &   $1{\vert}laminar$, $non{\verb'-'}pmtn{\vert}E$   & \multicolumn{1}{c|}{}  &   NP{\verb'-'}hard \cite{AntoniadisScheduling13}   &   QPTAS \cite{HuangMFCS14}   \\%quasipolynomial{\verb'-'}time approximation scheme
\multicolumn{1}{c|}{}   &   \multicolumn{1}{c|}{}   &   $1{\vert}r_j$, $d_j$, $non{\verb'-'}pmtn{\vert}E$   &  \multicolumn{1}{c|}{} &   NP{\verb'-'}hard   &   $2^{5{\alpha}-4}$ \cite{AntoniadisScheduling13}   \\
\multicolumn{1}{c|}{}   &   \multicolumn{1}{c|}{}   &      & \multicolumn{1}{c|}{}  &      &   $(1+\frac{w_{max}}{w_{min}})^{\alpha}$ \cite{BampisDAM15}   \\
\multicolumn{1}{c|}{}   &   \multicolumn{1}{c|}{}   &      & \multicolumn{1}{c|}{}  &      &   $2^{{\alpha}-1}(1+\epsilon)^{\alpha}\widetilde{B}_{\alpha}$ \cite{BampisFSTTCS13}   \\
\multicolumn{1}{c|}{}   &   \multicolumn{1}{c|}{}   &      &  \multicolumn{1}{c|}{} &      &   $(12(1+\epsilon))^{\alpha-1}$ \cite{Cohen-AddadWAOA14}   \\
\cline{2-6}
\multicolumn{1}{c|}{}   &   \multicolumn{1}{c|}{\multirow{16}{2.2cm}{Multiple homogeneous processors}}   &   $P{\vert}r_j=0$, $d_j=d$, $pmtn$, $mig{\vert}E$   &  \multicolumn{1}{c|}{\multirow{3}{2.5cm}{Offline}} &   $O(n\log{n})$ \cite{ChenECTRS04}   &  \verb'--'  \\
\multicolumn{1}{c|}{}   &   \multicolumn{1}{c|}{}   &   $P{\vert}r_j$, $d_j$, $pmtn$, $mig{\vert}E$   & \multicolumn{1}{c|}{}  &$O(n^2f(n))$ \cite{AlbersSPAA11}&   \verb'--'  \\
\multicolumn{1}{c|}{}   &   \multicolumn{1}{c|}{}   &      &  \multicolumn{1}{c|}{} &$O(nf(n)\log{K})$ \cite{EngelEuroPar12}&   \verb'--'  \\
\cline{3-6}
\multicolumn{1}{c|}{}   &   \multicolumn{1}{c|}{}   &   $P{\vert}agreeable$, $w_j=1$, $pmtn$, $non{\verb'-'}mig{\vert}E$   & \multicolumn{1}{c|}{\multirow{1}{2.5cm}{Offline}}  & $O(n\log{n})$  \cite{AlbersSPAA07} &   \verb'--'  \\
\cline{4-6}
\multicolumn{1}{c|}{}   &   \multicolumn{1}{c|}{}   &      & \multicolumn{1}{c|}{\multirow{2}{2.5cm}{Online}}  &  &   $2(\frac{\alpha}{\alpha-1})^{\alpha}e^{\alpha}$  \cite{AlbersSPAA07}  \\
\multicolumn{1}{c|}{}   &   \multicolumn{1}{c|}{}   &   $P{\vert}agreeable$, $pmtn$, $non{\verb'-'}mig{\vert}E$     & \multicolumn{1}{c|}{}  &  &   ${\alpha}^{\alpha}2^{4\alpha}$  \cite{AlbersSPAA07}  \\
\cline{4-6}
\multicolumn{1}{c|}{}   &   \multicolumn{1}{c|}{}   &   $P{\vert}r_j=0$, $d_j=d$, $w_j=1$, $\mathcal{M}_j$, $pmtn$, $non{\verb'-'}mig{\vert}E$   & \multicolumn{1}{c|}{\multirow{3}{2.3cm}{Offline}}  &  $O(mf(n)\log{n})$  \cite{JinEuroPar13} &   \verb'--'  \\
\multicolumn{1}{c|}{}   &   \multicolumn{1}{c|}{}   &   $P{\vert}r_j=0$, $d_j=d$, $pmtn$, $non{\verb'-'}mig{\vert}E$   & \multicolumn{1}{c|}{}  &   NP{\verb'-'}hard \cite{JinEuroPar13, AlbersSPAA07}   &   PTAS \cite{JinEuroPar13, AlbersSPAA07}   \\
\multicolumn{1}{c|}{}   &   \multicolumn{1}{c|}{}   &   $P{\vert}r_j$, $d_j$, $w_j=1$, $pmtn$, $non{\verb'-'}mig{\vert}E$   &   \multicolumn{1}{c|}{}   &   NP{\verb'-'}hard $(m{\geq}2)$ \cite{AlbersSPAA07}   &   $\alpha^{\alpha}2^{4\alpha}$ \cite{AlbersSPAA07}, $B_{{\lceil}\alpha{\rceil}}$ \cite{GreinerSPAA09}   \\
\cline{4-6}
\multicolumn{1}{c|}{}   &   \multicolumn{1}{c|}{}   &   & \multicolumn{1}{c|}{\multirow{1}{2.3cm}{Online}}  &      &   $\alpha^{\alpha}2^{4\alpha}$ \cite{AlbersSPAA07}   \\
\cline{4-6}
\multicolumn{1}{c|}{}   &   \multicolumn{1}{c|}{}   &   $P{\vert}r_j=0$, $d_j$, $pmtn$, $non{\verb'-'}mig{\vert}E$   &  \multicolumn{1}{c|}{\multirow{2}{2.3cm}{Offline}} &   NP{\verb'-'}hard   &$2(2-\frac{1}{m})^{\alpha}$\cite{AlbersSPAA07}, $B_{{\lceil}\alpha{\rceil}}$\cite{GreinerSPAA09}\\
\multicolumn{1}{c|}{}   &   \multicolumn{1}{c|}{}   &   $P{\vert}r_j$, $d_j$, $pmtn$, $non{\verb'-'}mig{\vert}E$   & \multicolumn{1}{c|}{}  &   NP{\verb'-'}hard   &   $B_{{\lceil}\alpha{\rceil}}$ \cite{GreinerSPAA09}   \\
\cline{4-6}
\multicolumn{1}{c|}{}   &   \multicolumn{1}{c|}{}   &     & \multicolumn{1}{c|}{\multirow{1}{2.3cm}{Online}}  &      &   $2(\frac{\alpha}{\alpha-1})^{\alpha}e^{\alpha}B_{{\lceil}\alpha{\rceil}}$ \cite{GreinerSPAA09}   \\
\cline{4-6}
\multicolumn{1}{c|}{}   &   \multicolumn{1}{c|}{}   &   $P{\vert}r_j=0$, $d_j=1$, $w_{i,j}$, $pmtn$, $non{\verb'-'}mig{\vert}E$   & \multicolumn{1}{c|}{\multirow{1}{2.3cm}{Offline}}  &   APX{\verb'-'}hard \cite{Cohen-AddadWAOA14}   &   \\
\cline{3-6}
\multicolumn{1}{c|}{}   &   \multicolumn{1}{c|}{}   &   $P{\vert}r_j$, $d_j$, $non{\verb'-'}pmtn$, $non{\verb'-'}mig{\vert}E$   &  \multicolumn{1}{c|}{\multirow{2}{2.3cm}{Offline}}  &   NP{\verb'-'}hard  &   $m^{\alpha}(n^{\frac{1}{m}})^{\alpha-1}$ \cite{BampisDAM15}   \\
\multicolumn{1}{c|}{}   &   \multicolumn{1}{c|}{}   &   & \multicolumn{1}{c|}{}  &  &   $({\frac{5}{2}})^{\alpha-1}{\widetilde{B}_{\alpha}}((1+\epsilon)(1+\frac{w_{max}}{w_{min}}))^{\alpha}$\cite{Cohen-AddadWAOA14}   \\
\cline{2-6}
\multicolumn{1}{c|}{}   &   \multicolumn{1}{c|}{\multirow{2}{2.2cm}{Multiple heterogeneous processors}}   &   $R{\vert}r_{ij}$, $d_{ij}$, $w_{ij}$, $pmtn$, $mig{\vert}E$   & \multicolumn{1}{c|}{\multirow{2}{2.3cm}{Offline}}  &   Polynomial $(\frac{1}{\epsilon})$   &  $OPT+\epsilon$ \cite{BampisFSTTCS13}  \\
\multicolumn{1}{c|}{}   &   \multicolumn{1}{c|}{}   &   $R{\vert}r_{ij}$, $d_{ij}$, $w_{ij}$, $pmtn$, $non{\verb'-'}mig{\vert}E$   &  \multicolumn{1}{c|}{}  &   NP{\verb'-'}hard   &  $(1+\epsilon)^{\alpha}{\widetilde{B}_{\alpha}}$  \cite{BampisFSTTCS13}  \\
\hline
\multicolumn{1}{c|}{\multirow{5}{1.6cm}{Discrete speed scaling}}   &   \multicolumn{1}{c|}{\multirow{2}{2.2cm}{Single processor}}   &   $1{\vert}r_j$, $d_j$, $pmtn{\vert}E$   & \multicolumn{1}{c|}{\multirow{1}{2.3cm}{Offline}}  &$O(dn\log{n})$ \cite{LiJSIAM05}& \verb'--'  \\
\cline{4-6}
\multicolumn{1}{c|}{}   &   \multicolumn{1}{c|}{}   &      & \multicolumn{1}{c|}{\multirow{1}{2.3cm}{Online}}  &   & $\frac{2^{\alpha-1}(\alpha-1)^{\alpha-1}(\delta^{\alpha}-1)^{\alpha}}{(\delta-1)(\delta^{\alpha}-\delta)^{\alpha-1}}+1$ ($\delta=max_{i}\frac{s_i}{s_{i+1}}$) \cite{LiTCS11}  \\ %O(\delta)
\cline{2-6}
\multicolumn{1}{c|}{}   &   \multicolumn{1}{c|}{\multirow{3}{2.2cm}{Multiple homogeneous processors}}   &   \multicolumn{1}{c|}{\multirow{3}{5.6cm}{$P{\vert}r_j=0$, $d_j=d$, $\mathcal{M}_j$, $pmtn$, $non{-}mig{\vert}E$}}   & \multicolumn{1}{c|}{\multirow{3}{2.3cm}{Offline}}  & \multicolumn{1}{c|}{\multirow{3}{2.6cm}{NP{-}hard \cite{JinIPDPSForum13}}} & \multicolumn{1}{c}{\multirow{3}{5.5cm}{$(2\Delta)^{\alpha-1}(\Delta=max_{i}(s_{i+1}/s_i))$ \cite{JinIPDPSForum13}}}  \\%{\verb'-'}
\multicolumn{1}{c|}{}   &   \multicolumn{1}{c|}{}   &   \multicolumn{1}{c|}{}   & \multicolumn{1}{c|}{}  & \multicolumn{1}{c|}{} &  \multicolumn{1}{c}{} \\
\multicolumn{1}{c|}{}   &   \multicolumn{1}{c|}{}   &   \multicolumn{1}{c|}{}   & \multicolumn{1}{c|}{}  & \multicolumn{1}{c|}{} &  \multicolumn{1}{c}{} \\
\cline{1-6}
\hline\hline
\end{tabular}
\label{tb:Algorithmic results of minimizing energy subject to deadline feasibility constraints on speed scaling}
\end{sidewaystable*}

\begin{sidewaystable*}[htbp]
\footnotesize
\renewcommand{\arraystretch}{1.3}
\centering
\caption{Algorithmic results for speed scaling approaches based on flow time, makespan, and throughput.}
\begin{tabular}{c|c|c|p{1.2cm}|c|c}
\hline\hline
\cline{1-6}
\multicolumn{1}{c|}{\multirow{2}{1.8cm}{Speed Mode}} & \multicolumn{1}{c|}{\multirow{2}{2.2cm}{Environment}} &   \multicolumn{1}{c|}{\multirow{2}{1.8cm}{Problem}}   & \multicolumn{1}{c|}{\multirow{2}{1.2cm}{Algorithm Type}} &   \multicolumn{1}{c|}{\multirow{2}{1.8cm}{Complexity}} &   \multicolumn{1}{c}{\multirow{2}{4cm}{Approximation/Competitive Ratio}} \\
\multicolumn{1}{c|}{}   &   \multicolumn{1}{c|}{}   &   \multicolumn{1}{c|}{}   & \multicolumn{1}{c|}{}  & \multicolumn{1}{c|}{} &  \multicolumn{1}{c}{} \\
\hline
\multicolumn{1}{c|}{\multirow{32}{1.8cm}{Continuous speed scaling}}   &   \multicolumn{1}{c|}{\multirow{13}{2.2cm}{Single processor}}   &   $1{\vert}r_j$, $w_j=1$, $non{\verb'-'}pmtn$, $E{\vert}F$   & \multicolumn{1}{c|}{\multirow{3}{0.8cm}{Offline}}  &$O(n^2\log{L})$ \cite{PruhsSWAT04}   & \verb'--'  \\
\multicolumn{1}{c|}{}   &   \multicolumn{1}{c|}{}   & $1{\vert}r_j$, $non{\verb'-'}pmtn$, $E{\vert}F$  & \multicolumn{1}{c|}{} &    & $(1+\epsilon)^{\alpha}${\verb'-'}energy $O(\frac{1}{\epsilon})${\verb'-'}approximate \cite{PruhsSWAT04}  \\
\multicolumn{1}{c|}{}   &   \multicolumn{1}{c|}{}   &  $1{\vert}r_j$, $w_j=1$, $non{\verb'-'}pmtn{\vert}E+F$  &  \multicolumn{1}{c|}{}   &  $O(n^3\log{K})$  \cite{AlbersATLG07} & \verb'--'  \\
\cline{4-6}
\multicolumn{1}{c|}{}   &   \multicolumn{1}{c|}{}   &   \multicolumn{1}{c|}{}   &  \multicolumn{1}{c|}{\multirow{1}{0.8cm}{Online}}  &  & $8.3e((3+\sqrt{5})/2)^{\alpha}$ \cite{AlbersATLG07}   \\
\cline{3-6}
\multicolumn{1}{c|}{}   &   \multicolumn{1}{c|}{}   &   $1{\vert}r_j$, $w_j=1$, $pmtn{\vert}E+F$   &   \multicolumn{1}{c|}{\multirow{6}{0.8cm}{Online}}   &   &  4 \cite{BansalSODA07}  \\
\multicolumn{1}{c|}{}   &   \multicolumn{1}{c|}{}   &   $1{\vert}r_j$, $pmtn{\vert}E+WF$   & \multicolumn{1}{c|}{}  &   &  $O({\alpha}^2/\ln^2{\alpha})$ \cite{BansalSODA07}  \\
\multicolumn{1}{c|}{}   &   \multicolumn{1}{c|}{}   &   $1{\vert}r_j$, $pmtn$, $E{\vert}F$   & \multicolumn{1}{c|}{}  &   &  not $O(1)$-competitive \cite{BansalSODA07}  \\
\multicolumn{1}{c|}{}   &   \multicolumn{1}{c|}{}   &   $1{\vert}r_j$, $pmtn{\vert}E+F$   & \multicolumn{1}{c|}{}  &   &  $O(\frac{\alpha}{\ln{\alpha}})$ \cite{LamESA08}  \\  
\multicolumn{1}{c|}{}   &   \multicolumn{1}{c|}{}   &   \multicolumn{1}{c|}{}   & \multicolumn{1}{c|}{}  &   &  $3+\epsilon$ \cite{BansalSODA09} \\
\multicolumn{1}{c|}{}   &   \multicolumn{1}{c|}{}   &   \multicolumn{1}{c|}{}   & \multicolumn{1}{c|}{}  &   &  $2+\epsilon$ \cite{AndrewSIGMETRICS09} \\
\cline{4-6}
\multicolumn{1}{c|}{}   &   \multicolumn{1}{c|}{}   &   $1{\vert}r_j$, $pmtn/non{\verb'-'}pmtn$, $E{\vert}C_{max}$   & \multicolumn{1}{c|}{\multirow{3}{0.8cm}{Offline}}   & $O(n\log{n})$ \cite{BundeScheduling09}  &  \verb'--'  \\
\multicolumn{1}{c|}{}   &   \multicolumn{1}{c|}{}   &   $1{\vert}r_j$, $d_j$, $pmtn$, $E{\vert}U$   & \multicolumn{1}{c|}{}   & Pseudo-poly, $O(n^6T^9V_{max}^9)$ \cite{AngelSTACS14}  &  \verb'--'  \\
\multicolumn{1}{c|}{}   &   \multicolumn{1}{c|}{}   &   $1{\vert}r_j$, $d_j$, $pmtn$, $E{\vert}WU$   & \multicolumn{1}{c|}{}   & Pseudo-poly, $O(n^2S^4T^9V_{max}^9)$ \cite{AngelSTACS14}  &  \verb'--' \nocite{aaaa}  \\
\cline{2-6}
\multicolumn{1}{c|}{}   &   \multicolumn{1}{c|}{\multirow{5}{2.2cm}{Single processor (speed bounded)}}   &   $1{\vert}r_j$, $w_j=1$, $pmtn{\vert}E+F$   & \multicolumn{1}{c|}{\multirow{4}{0.8cm}{Online}}  &   & 4 \cite{BansalICALP08}  \\
\multicolumn{1}{c|}{}   &   \multicolumn{1}{c|}{}   &   $1{\vert}r_j$, $pmtn{\vert}E+F$   & \multicolumn{1}{c|}{}  &   & ${2(\alpha+1)}/{(\alpha-\frac{\alpha-1}{(\alpha+1)^{1/(\alpha-1)}})}$ \cite{LamESA08}  \\
\multicolumn{1}{c|}{}   &   \multicolumn{1}{c|}{}   &   $1{\vert}r_j$, $d_j$, $pmtn$, ${\vert}U, E$   & \multicolumn{1}{c|}{}   &   & $14$-competitive $(\alpha^{\alpha}+\alpha^24^{\alpha})$-competitive \cite{ChanSODA07}   \\
\multicolumn{1}{c|}{}   &   \multicolumn{1}{c|}{}   &      & \multicolumn{1}{c|}{}   &   & $4$-competitive $(\alpha^{\alpha}+\alpha^24^{\alpha})$-competitive \cite{BansalICALP08}   \\
\cline{4-6}
\multicolumn{1}{c|}{}   &   \multicolumn{1}{c|}{}   &      & \multicolumn{1}{c|}{\multirow{1}{0.8cm}{Offline}}   &   & $3$-approximate $O(1)$-approximate \cite{LiTCS11}   \\

\cline{2-6}
\multicolumn{1}{c|}{}   &   \multicolumn{1}{c|}{\multirow{2}{2.2cm}{Multiple homogeneous processors (speed bounded)}}   &   $P=2{\vert}r_j$, $d_j$, $pmtn$, $mig{\vert}U, E$   &  \multicolumn{1}{c|}{\multirow{3}{0.8cm}{Online}} &    &  $3$-competitive $O(1)$-competitive \cite{LamISAAC07}  \\
\multicolumn{1}{c|}{}   &   \multicolumn{1}{c|}{}   &   $P{\vert}r_j$, $pmtn$, $mig{\vert}E+F$   &  \multicolumn{1}{c|}{} &    &  $O(1)$ \cite{LamScheduling12}  \\
\multicolumn{1}{c|}{}   &   \multicolumn{1}{c|}{}   &   $P{\vert}r_j$, $pmtn$, $non{\verb'-'}mig{\vert}E+F$   &  \multicolumn{1}{c|}{} &    &  $O(1)$ \cite{LamScheduling12}  \\
\cline{2-6}
\multicolumn{1}{c|}{}   &   \multicolumn{1}{c|}{\multirow{7}{2.2cm}{Multiple homogeneous processors}}   &   $P{\vert}r_j$, $w_j=1$, $pmtn/non{\verb'-'}pmtn$, $non{\verb'-'}mig$, $E{\vert}F$   &  \multicolumn{1}{c|}{\multirow{1}{0.8cm}{Offline}} &    &  arbitrarily good \cite{BundeScheduling09} \\
\cline{4-6}
\multicolumn{1}{c|}{}   &   \multicolumn{1}{c|}{}   &   $P{\vert}r_j$, $pmtn$, $non{\verb'-'}mig{\vert}E+F$   & \multicolumn{1}{c|}{\multirow{1}{0.8cm}{Online}}  &   &  $(3+\epsilon)B_{{\lceil}\alpha{\rceil}}$ \cite{GreinerSPAA09} \\
\cline{4-6}
\multicolumn{1}{c|}{}   &   \multicolumn{1}{c|}{}   &   $P{\vert}r_j$, $w_j=1$, $pmtn/non{\verb'-'}pmtn$, $non{\verb'-'}mig$, $E{\vert}C_{max}$   &  \multicolumn{1}{c|}{\multirow{5}{0.8cm}{Offline}} &    &  arbitrarily good \cite{BundeScheduling09} \\
\multicolumn{1}{c|}{}   &   \multicolumn{1}{c|}{}   &   $P{\vert}r_j=0$, $non{\verb'-'}pmtn$, $non{\verb'-'}mig$, $E{\vert}C_{max}$   &  \multicolumn{1}{c|}{} &  NP{\verb'-'}hard \cite{BundeScheduling09}  &   \\
\multicolumn{1}{c|}{}   &   \multicolumn{1}{c|}{}   &   $P{\vert}r_j=0$, $non{\verb'-'}pmtn$, $non{\verb'-'}mig$, $prec$, $E{\vert}C_{max}$   &  \multicolumn{1}{c|}{} &  NP{\verb'-'}hard  &  $O(\log^{1+\frac{2}{\alpha}}m)$ \cite{PruhsTCS08} \\
\multicolumn{1}{c|}{}   &   \multicolumn{1}{c|}{}   &   $P{\vert}r_j$, $d_j$, $w_j=1$, $m$, $non{\verb'-'}pmtn$, $non{\verb'-'}mig$, $E{\vert}WU$   &  \multicolumn{1}{c|}{} & pseudo, $O(n^{12m+7}S^2)$ \cite{AngelISAAC14}  &   \\
\multicolumn{1}{c|}{}   &   \multicolumn{1}{c|}{}   &   $P{\vert}agreeable$, $w_j$, $m$, $non{\verb'-'}pmtn$, $non{\verb'-'}mig$, $E{\vert}WU$   &  \multicolumn{1}{c|}{} & $O(n^{2m+2}V^{2m+1}Sm)$ \cite{AngelISAAC14}  &   \\
\cline{2-6}
\multicolumn{1}{c|}{}   &   \multicolumn{1}{c|}{\multirow{4}{2.2cm}{Multiple heterogeneous processors}}   &   $R{\vert}r_{j}$, $pmtn$, $non{\verb'-'}mig{\vert}E+F$   & \multicolumn{1}{c|}{\multirow{3}{0.8cm}{Online}}  &     &  $O(\frac{1}{\epsilon})$-competitive $(1+\epsilon)$-speedup  \cite{GuptaICALP10} \\
\multicolumn{1}{c|}{}   &   \multicolumn{1}{c|}{}   &   $R{\vert}r_{j}$, $P_i(s)=s^{\alpha_i}$, $pmtn$, $non{\verb'-'}mig{\vert}E+WF$   &  \multicolumn{1}{c|}{}  &      &  $O(\alpha^2)$ $(\alpha=max_i\alpha_i)$  \cite{GuptaICALP10}  \\
\multicolumn{1}{c|}{}   &   \multicolumn{1}{c|}{}   &   $R{\vert}r_{j}$, $P_i(s)=s^{\alpha_i}$, $pmtn$, $non{\verb'-'}mig{\vert}E+F$   &  \multicolumn{1}{c|}{}  &      &  $O(\alpha)$ $(\alpha=max_i\alpha_i)$  \cite{GuptaICALP10}  \\
\cline{4-6}
\multicolumn{1}{c|}{}   &   \multicolumn{1}{c|}{}   &   $R{\vert}r_{ij}$, $d_{ij}$, $w_{ij}$, $P_i(s)=s^{\alpha_i}$, $pmtn$, $non{\verb'-'}mig$, $E{\vert}WU$   &  \multicolumn{1}{c|}{\multirow{1}{0.8cm}{Offline}}  &      &  $2(\Gamma+1)(1+\epsilon)$ $(\Gamma=max_i\alpha_i)$  \cite{AngelISAAC14}  \\
\hline
\multicolumn{1}{c|}{\multirow{2}{1.8cm}{Discrete speed scaling}}   &   \multicolumn{1}{c|}{\multirow{2}{2.2cm}{Single processor}}   &   $1{\vert}r_j$, $pmtn/non{\verb'-'}pmtn$, $E{\vert}C_{max}$   & \multicolumn{1}{c|}{\multirow{1}{0.8cm}{Offline}}   & $O(n\log{n})$ \cite{BundeScheduling09}  &  \verb'--'  \\
\cline{4-6}
\multicolumn{1}{c|}{}   &  \multicolumn{1}{c|}{}    &   $1{\vert}r_j$, $d_j$, $pmtn{\vert}U, E$   & \multicolumn{1}{c|}{\multirow{1}{0.8cm}{Online}}  & & $14$-competitive $(\Delta^{\alpha}(\alpha^{\alpha}+\alpha^{\alpha}4^{\alpha})+1)$-competitive \cite{ChanSODA07} \\
\cline{1-6}
\hline\hline
\end{tabular}
\label{tb:Algorithmic results of flow time, makespan, throughput on speed scaling}
\end{sidewaystable*}

\renewcommand{\arraystretch}{1.3}
\begin{table}[htbp]
\centering
\footnotesize
\caption{\label{tb:Definitions of symbol on dynamic speed scaling} Definitions of symbols for dynamic speed scaling}
\begin{tabular}{p{1.45cm}|p{6.4cm}}
\hline\hline
Symbol & Definition  \\\hline
$1$&Single process\\
$P$&Homogeneous parallel (multiple) processors\\
$R$&Heterogeneous parallel (multiple) processors\\
$m$&Number of processors\\
$n$&Number of jobs\\
$d$&Number of available discrete speeds\\
\hline
$e$&A mathematical constant, $e{\simeq}2.71828$\\
$\alpha$&A constant power parameter\\
$\epsilon$&An arbitrary parameter\\
$B_{{\lceil}\alpha{\rceil}}$&${\lceil}\alpha{\rceil}$-th Bell number\\
${\widetilde{B}_{\alpha}}$&Generalized version of the Bell number $\Omega((\frac{\alpha}{e\ln{\alpha}})^{\alpha})$\\
$\Delta$&The largest ratio of two consecutive (non-zero) speed levels\\
\hline
\multicolumn{1}{c|}{\multirow{2}{1.45cm}{$f(n)$}}&\multicolumn{1}{c}{\multirow{2}{6.4cm}{Time complexity of computing a maximum flow in a layered graph with $n$ vertices}}\\
\multicolumn{1}{c|}{}   &   \multicolumn{1}{c}{}\\
\hline
\multicolumn{1}{c|}{\multirow{2}{1.45cm}{$K$}}&\multicolumn{1}{c}{\multirow{2}{6.4cm}{The range of all possible values of speeds divided by the desired accuracy}}\\
\multicolumn{1}{c|}{}   &   \multicolumn{1}{c}{}\\
\hline
\multicolumn{1}{c|}{\multirow{2}{1.45cm}{$L$}}&\multicolumn{1}{c}{\multirow{2}{6.4cm}{The range of all possible values of power divided by the desired accuracy}}\\
\multicolumn{1}{c|}{}   &   \multicolumn{1}{c}{}\\
\hline
$T$&The span length $max_{j,j'}(d_j-r_{j'})$\\
$V_{max}$&The maximum processing volume\\
$V$&The sum of processing volumes $\Sigma_jw_j$\\
$S$&The sum of weights\\ 
\hline
\hline
$r_j$&Job $j$'s release date\\
$d_j$&Job $j$'s deadline\\
$w_j$&Job $j$'s processing volume\\
$w_{max}(w_{min})$&Maximum (Minimum) processing volume of all jobs\\
$r_{ij}$&Job $j$'s release date on processor $i$\\
$d_{ij}$&Job $j$'s deadline on processor $i$\\
$w_{ij}$&Job $j$'s processing volume on processor $i$\\
$\mathcal{M}_j$&Job $j$'s eligible processor set\\
$E$&Energy consumption\\
$(W)F$&(Weight) Flow time\\
$C_{max}$&Makespan\\
$(W)U$&(Weight) Throughput\\
$pre$&Precedence constraints defined between jobs\\
$(non-)pmtn$&(Not) Allow jobs to be interrupted\\
\hline
\multicolumn{1}{c|}{\multirow{2}{1.45cm}{$(non-)mig$}}&\multicolumn{1}{c}{\multirow{2}{6.4cm}{(Not) Allow jobs to be interrupted and resumed on the same or another processor}}\\
\multicolumn{1}{c|}{}   &   \multicolumn{1}{c}{}\\
\hline
$agreeable$&For any two jobs $j$ and $j'$, if $r_j{\leq}r_{j'}$, then $d_j{\leq}d_{j'}$\\
$laminar$&If $r_j{\leq}r_{j'}$, then $d_j{\geq}d_{j'}$ or $d_j{\leq}r_{j'}$\\
\hline\hline
\end{tabular}
\end{table}

When a large-scale warehouse is considered, speed scaling is adopted to seek solutions for optimizing the energy consumption based on the current load \cite{GandhiSIGMETRICS09, WiermanINFOCOM09, LiuTPDS14}. Higher speeds allow for faster execution but also result in higher power consumption. Gandhi \emph{et al.} \cite{GandhiSIGMETRICS09} studied the problem of finding the optimal power allocating in terms of determining the optimal frequencies of the servers in a server farm to minimize mean response time after measuring the power-to-frequency relationship of each server for a given workload. The authors noted that use of the optimal power allocation significantly improved the response time, by a factor of typically 1.4 and, in some cases, as much as a factor of 5. In \cite{WiermanINFOCOM09}, the authors considered how to balance the mean response time and mean energy consumption in processor sharing scheduling when dynamic speed scaling is applied to reduce energy consumption. They identified a scheme that provided results that were nearly identical to the optimal objective of dynamic speed scaling, i.e., that simultaneously minimized energy consumption and response time. Speed scaling has also been studied in the arbitration of power consumption and system throughput. In \cite{LiuTPDS14}, the authors provided a probabilistic framework in which online decisions were made on request for admission control, routing, and virtual machine (VM) allocation. They modeled a unified objective function to balance the power consumption of the servers and the system throughput for application requests.

At the network level, an Adaptive Link Rate (ALR) can be applied on links to reduce energy consumption \cite{BilalCluster13, GunaratneTOC08, AbtsISCA10, WangNCA13}. Gunaratne \emph{et al.} \cite{GunaratneTOC08} proposed a Markovian model to reduce the energy consumption of a typical Ethernet link by adaptively varying the link data rate in response to utilization. The authors found that their scheme allows an Ethernet link to operate at a lower data rate for more than 80\% of the time, yielding approximately {\$}70 million in energy savings per year in the U.S. with only a very small increase in packet delay. Abts \emph{et al.} \cite{AbtsISCA10} proposed a power-optimizing mechanism for links with the capabilities of a network with the flattened butterfly topology. The authors scaled down the network links with a data rate proportional to the traffic intensity while accepting a trade-off of additional mean latency. They reported an energy savings of approximately 42\%. Wang \emph{et al.} \cite{WangNCA13} presented a rate-adaptation-based solution with the intent of approaching network-wide energy proportionality via routing optimization. The simulation results indicated that the scheme could achieve up to 40\% energy savings while introducing a very slight increase in network delay.
\subsection{Power-Down Mechanism (PDM)}
In this scheme, devices are transitioned into a low-power standby or sleep mode while they are idle to save energy. One must determine when idle periods exist that are of sufficient duration to outweigh the costs incurred by the scheme (such as the cost for transitioning from the sleep state and the delay cost) and decide when to wake the devices from the power-saving mode to satisfy the workload demand. When a single device is considered and the transition delay is ignored, the problem of managing power in a two-state situation (active state and sleep state) is equivalent to the ski-rental problem. A competition analysis of this problem was presented by \cite{IraniHochbaum97, KarlinALGO94}. Irani \emph{et al.} \cite{IraniTECS03} extended the study to devices with multiple states. The authors presented a deterministic algorithm that achieved 2-competitive performance when the transition energies were assumed to be additive. In \cite{AugustineSIAMJ08}, the authors investigated the generalized case in which the state transition energies could take arbitrary values and achieved a competitiveness of $3+2\sqrt{2}$. With preemptions allowed and the process of waking up assumed to require a certain fixed amount of energy, Baptiste \emph{et al.} \cite{BaptisteESA07} provided an $O(n^5)$-time algorithm for offline operation. For the case in which the tasks have agreeable deadlines, i.e., later-released tasks also have later deadlines, \cite{GururajJalanStein10} and \cite{AngelDiscreteAppliedMathematics14} proposed algorithms to improve the time complexity. Given a single processor, tasks with arbitrary processing time and unit transition costs, \cite{GururajJalanStein10} proposed an $O(n\log{n})$ greedy algorithm. Later, Angel \emph{et al.} \cite{AngelDiscreteAppliedMathematics14} proposed an $O(n^2)$ algorithm for tasks with arbitrary processing time and arbitrary transition costs in a single-processor environment. For tasks with unit processing time and unit transition cost in the multiprocessor case, the authors of \cite{AngelDiscreteAppliedMathematics14} also provided an $O(n^2m)$ algorithm. In \cite{DemaineScheduling13}, Demaine \emph{et al.} considered the power-saving problem in which each task must be executed within a specified set of time intervals and proposed a $(1+\frac{2}{3}L_c)$-approximation algorithm, where $L_c$ is the transition cost. In \cite{ChenJJSTACS14}, the authors studied the problem of online dynamic power management in a three-state situation (busy state, standby state and off state). They first considered the case in which the set of tasks that could feasibly be scheduled on a single processor was known in advance (packable) and, for this case, proposed a $4$-competitive online algorithm that used at most two processors. The competitive ratio was improved to $3.59$ for tasks with unit processing time. For the general case of tasks with unit processing time and multiple processors, an $O(1)$-competitive algorithm was provided. In Table~\ref{tb:Algorithmic results regarding power-down mechanism}, we summarize all the algorithmic results mentioned above regarding power-down mechanisms; the definitions of the symbols are the same as those given in Table~\ref{tb:Definitions of symbol on dynamic speed scaling}.
\renewcommand{\arraystretch}{1.3}
\begin{table*}[htbp]
\centering
\scriptsize
\caption{\label{tb:Algorithmic results regarding power-down mechanism} Algorithmic results for power-down mechanisms. The definitions of the symbols are the same as those given in Table~\ref{tb:Definitions of symbol on dynamic speed scaling}, and the term $packable$ means that the set of tasks that can feasibly be scheduled on a single processor is known in advance.}
\begin{tabular}{p{1.8cm}|p{2.5cm}|p{4.4cm}|p{1.8cm}|p{1.5cm}|p{3.0cm}}
\hline\hline
\multicolumn{1}{c|}{\multirow{2}{1.8cm}{Environment}} & \multicolumn{1}{c|}{\multirow{2}{2.5cm}{Task Model}} & \multicolumn{1}{c|}{\multirow{2}{4.4cm}{Transition Cost}} & \multicolumn{1}{c|}{\multirow{2}{1.8cm}{Algorithm Type}} & \multicolumn{1}{c|}{\multirow{2}{1.5cm}{Complexity}} & \multicolumn{1}{c}{\multirow{2}{3.0cm}{Approximation/Competitive Ratio}} \\
\multicolumn{1}{c|}{}   &   \multicolumn{1}{c|}{}   &   \multicolumn{1}{c|}{}   & \multicolumn{1}{c|}{}  & \multicolumn{1}{c|}{} &  \multicolumn{1}{c}{} \\\hline
\multicolumn{1}{c|}{\multirow{16}{1.8cm}{Single processor}} & \multicolumn{1}{c|}{\multirow{2}{2.0cm}{$r_j$, $w_j$}} & \multicolumn{1}{c|}{\multirow{2}{4.4cm}{Two states (active state and sleep state), transition cost $L_c$ is arbitrary}} & \multicolumn{1}{c|}{\multirow{2}{1.8cm}{Online}} & \multicolumn{1}{c|}{\multirow{2}{1.8cm}{-\,-}} & \multicolumn{1}{c}{\multirow{2}{3.0cm}{$2$ \cite{IraniHochbaum97}, $e/(e-1)$ \cite{KarlinALGO94}}} \\
\multicolumn{1}{c|}{}   &   \multicolumn{1}{c|}{}   &   \multicolumn{1}{c|}{}   & \multicolumn{1}{c|}{}  & \multicolumn{1}{c|}{} &  \multicolumn{1}{c}{} \\
\cline{2-6}
\multicolumn{1}{c|}{} & \multicolumn{1}{c|}{\multirow{2}{2.0cm}{$r_j$, $w_j$}} & \multicolumn{1}{c|}{\multirow{2}{4.4cm}{Multiple states ($\{s_0,...,s_k\}$), transition cost $L_c$ is additive}} & \multicolumn{1}{c|}{\multirow{2}{1.8cm}{Online}} & \multicolumn{1}{c|}{\multirow{2}{1.8cm}{-\,-}} & \multicolumn{1}{c}{\multirow{2}{1.8cm}{$2$ \cite{IraniTECS03}}} \\
\multicolumn{1}{c|}{}   &   \multicolumn{1}{c|}{}   &   \multicolumn{1}{c|}{}   & \multicolumn{1}{c|}{}  & \multicolumn{1}{c|}{} &  \multicolumn{1}{c}{} \\
\cline{2-6}
\multicolumn{1}{c|}{} & \multicolumn{1}{c|}{\multirow{2}{2.0cm}{$r_j$, $w_j$}} & \multicolumn{1}{c|}{\multirow{2}{4.4cm}{Multiple states ($\{s_0,...,s_k\}$), transition cost $L_c$ is arbitrary}} & \multicolumn{1}{c|}{\multirow{2}{1.8cm}{Online}} & \multicolumn{1}{c|}{\multirow{2}{1.8cm}{-\,-}} & \multicolumn{1}{c}{\multirow{2}{1.8cm}{$3+2\sqrt{2}$ \cite{AugustineSIAMJ08}}} \\
\multicolumn{1}{c|}{}   &   \multicolumn{1}{c|}{}   &   \multicolumn{1}{c|}{}   & \multicolumn{1}{c|}{}  & \multicolumn{1}{c|}{} &  \multicolumn{1}{c}{} \\
\cline{2-6}
\multicolumn{1}{c|}{} & \multicolumn{1}{c|}{\multirow{2}{2.0cm}{$r_j$, $d_j$, $w_j=1$}} & \multicolumn{1}{c|}{\multirow{2}{4.4cm}{Two states (active state and sleep state), transition cost $L_c$ is arbitrary}} & \multicolumn{1}{c|}{\multirow{2}{1.8cm}{Offline}} & \multicolumn{1}{c|}{\multirow{2}{1.8cm}{$O(n^7)$ \cite{BaptisteSODA06}, $O(n^4)$ \cite{BaptisteESA07}}} & \multicolumn{1}{c}{\multirow{2}{1.8cm}{-\,-}} \\
\multicolumn{1}{c|}{}   &   \multicolumn{1}{c|}{}   &   \multicolumn{1}{c|}{}   & \multicolumn{1}{c|}{}  & \multicolumn{1}{c|}{} &  \multicolumn{1}{c}{} \\
\cline{2-6}
\multicolumn{1}{c|}{} & \multicolumn{1}{c|}{\multirow{2}{2.0cm}{$r_j$, $d_j$, $w_j$, $pmtn$}} & \multicolumn{1}{c|}{\multirow{2}{4.4cm}{Two states (active state and sleep state), transition cost $L_c$ is arbitrary}} & \multicolumn{1}{c|}{\multirow{2}{1.8cm}{Offline}} & \multicolumn{1}{c|}{\multirow{2}{1.8cm}{$O(n^5)$ \cite{BaptisteESA07}}} & \multicolumn{1}{c}{\multirow{2}{1.8cm}{-\,-}} \\
\multicolumn{1}{c|}{}   &   \multicolumn{1}{c|}{}   &   \multicolumn{1}{c|}{}   & \multicolumn{1}{c|}{}  & \multicolumn{1}{c|}{} &  \multicolumn{1}{c}{} \\
\cline{2-6}
\multicolumn{1}{c|}{} & \multicolumn{1}{c|}{\multirow{2}{2.0cm}{$agreeable$, $w_j$, $pmtn$}} & \multicolumn{1}{c|}{\multirow{2}{4.4cm}{Two states (active state and sleep state), transition cost $L_c$ is unit}} & \multicolumn{1}{c|}{\multirow{2}{1.8cm}{Offline}} & \multicolumn{1}{c|}{\multirow{2}{1.8cm}{$O(n\log{n})$ \cite{GururajJalanStein10}}} & \multicolumn{1}{c}{\multirow{2}{1.8cm}{-\,-}} \\
\multicolumn{1}{c|}{}   &   \multicolumn{1}{c|}{}   &   \multicolumn{1}{c|}{}   & \multicolumn{1}{c|}{}  & \multicolumn{1}{c|}{} &  \multicolumn{1}{c}{} \\
\cline{2-6}
\multicolumn{1}{c|}{} & \multicolumn{1}{c|}{\multirow{2}{2.0cm}{$agreeable$, $w_j=1$}} & \multicolumn{1}{c|}{\multirow{2}{4.4cm}{Two states (active state and sleep state), transition cost $L_c$ is arbitrary}} & \multicolumn{1}{c|}{\multirow{2}{1.8cm}{Offline}} & \multicolumn{1}{c|}{\multirow{2}{1.8cm}{$O(n^3)$ \cite{GururajJalanStein10}, $O(n^2)$ \cite{AngelDiscreteAppliedMathematics14}}} & \multicolumn{1}{c}{\multirow{2}{1.8cm}{-\,-}} \\
\multicolumn{1}{c|}{}   &   \multicolumn{1}{c|}{}   &   \multicolumn{1}{c|}{}   & \multicolumn{1}{c|}{}  & \multicolumn{1}{c|}{} &  \multicolumn{1}{c}{} \\
\cline{2-6}
\multicolumn{1}{c|}{} & \multicolumn{1}{c|}{\multirow{2}{2.0cm}{$agreeable$, $w_j$, $pmtn$}} & \multicolumn{1}{c|}{\multirow{2}{4.4cm}{Two states (active state and sleep state), transition cost $L_c$ is arbitrary}} & \multicolumn{1}{c|}{\multirow{2}{1.8cm}{Offline}} & \multicolumn{1}{c|}{\multirow{2}{1.8cm}{$O(n^2)$ \cite{AngelDiscreteAppliedMathematics14}}} & \multicolumn{1}{c}{\multirow{2}{1.8cm}{-\,-}} \\
\multicolumn{1}{c|}{}   &   \multicolumn{1}{c|}{}   &   \multicolumn{1}{c|}{}   & \multicolumn{1}{c|}{}  & \multicolumn{1}{c|}{} &  \multicolumn{1}{c}{} \\
\hline
\multicolumn{1}{c|}{\multirow{7}{1.8cm}{Multiple homogeneous processors}} & \multicolumn{1}{c|}{\multirow{2}{2.0cm}{$agreeable$, $w_j=1$}} & \multicolumn{1}{c|}{\multirow{2}{4.4cm}{Two states (active state and sleep state), transition cost $L_c$ is unit}} & \multicolumn{1}{c|}{\multirow{2}{1.8cm}{Offline}} & \multicolumn{1}{c|}{\multirow{2}{1.8cm}{$O(m^2n^3)$ \cite{GururajJalanStein10}, $O(mn^2)$ \cite{AngelDiscreteAppliedMathematics14}}} & \multicolumn{1}{c}{\multirow{2}{1.8cm}{-\,-}} \\
\multicolumn{1}{c|}{}   &   \multicolumn{1}{c|}{}   &   \multicolumn{1}{c|}{}   & \multicolumn{1}{c|}{}  & \multicolumn{1}{c|}{} &  \multicolumn{1}{c}{} \\
\cline{2-6}
\multicolumn{1}{c|}{} & \multicolumn{1}{c|}{\multirow{2}{2.0cm}{$r_j$, $d_j$, $w_j=1$, $r_j{\in}\mathbb{N},d_j{\in}\mathbb{N}$}} & \multicolumn{1}{c|}{\multirow{2}{4.4cm}{Two states (active state and sleep state), transition cost $L_c$ is arbitrary}} & \multicolumn{1}{c|}{\multirow{2}{1.8cm}{Offline}} & \multicolumn{1}{c|}{\multirow{2}{1.8cm}{$O(m^5n^7)$ \cite{DemaineScheduling13}}} & \multicolumn{1}{c}{\multirow{2}{1.8cm}{-\,-}} \\
\multicolumn{1}{c|}{}   &   \multicolumn{1}{c|}{}   &   \multicolumn{1}{c|}{}   & \multicolumn{1}{c|}{}  & \multicolumn{1}{c|}{} &  \multicolumn{1}{c}{} \\
\cline{2-6}
\multicolumn{1}{c|}{} & \multicolumn{1}{c|}{\multirow{3}{2.5cm}{Execute in a specified set of time intervals, $w_j=1$}} & \multicolumn{1}{c|}{\multirow{3}{4.4cm}{Two states (active state and sleep state), transition cost $L_c$ is arbitrary}} & \multicolumn{1}{c|}{\multirow{3}{1.8cm}{Offline}} & \multicolumn{1}{c|}{\multirow{3}{1.8cm}{NP-hard \cite{DemaineScheduling13}}} & \multicolumn{1}{c}{\multirow{3}{1.8cm}{$1+\frac{2}{3}L_c$ \cite{DemaineScheduling13}}} \\
\multicolumn{1}{c|}{}   &   \multicolumn{1}{c|}{}   &   \multicolumn{1}{c|}{}   & \multicolumn{1}{c|}{}  & \multicolumn{1}{c|}{} &  \multicolumn{1}{c}{} \\
\multicolumn{1}{c|}{}   &   \multicolumn{1}{c|}{}   &   \multicolumn{1}{c|}{}   & \multicolumn{1}{c|}{}  & \multicolumn{1}{c|}{} &  \multicolumn{1}{c}{} \\
\cline{1-6}
\multicolumn{1}{c|}{\multirow{6}{1.8cm}{Multiple homogeneous processors, $m=2$}} & \multicolumn{1}{c|}{\multirow{3}{2.0cm}{$r_j$, $d_j$, $w_j=1$, $pmtn$, $non-mig$, $packable$}} & \multicolumn{1}{c|}{\multirow{3}{4.4cm}{Three states (busy, standby and off), transition cost $L_c$ is arbitrary}} & \multicolumn{1}{c|}{\multirow{3}{1.8cm}{Online}} & \multicolumn{1}{c|}{\multirow{3}{1.8cm}{-\,-}} & \multicolumn{1}{c}{\multirow{3}{1.8cm}{3.59 \cite{ChenJJSTACS14}}} \\
\multicolumn{1}{c|}{}   &   \multicolumn{1}{c|}{}   &   \multicolumn{1}{c|}{}   & \multicolumn{1}{c|}{}  & \multicolumn{1}{c|}{} &  \multicolumn{1}{c}{} \\
\multicolumn{1}{c|}{}   &   \multicolumn{1}{c|}{}   &   \multicolumn{1}{c|}{}   & \multicolumn{1}{c|}{}  & \multicolumn{1}{c|}{} &  \multicolumn{1}{c}{} \\
\cline{2-6}
\multicolumn{1}{c|}{} & \multicolumn{1}{c|}{\multirow{3}{2.0cm}{$r_j$, $d_j$, $w_j$, $pmtn$, $non-mig$, $packable$}} & \multicolumn{1}{c|}{\multirow{3}{4.4cm}{Three states (busy, standby and off), transition cost $L_c$ is arbitrary}} & \multicolumn{1}{c|}{\multirow{3}{1.8cm}{Online}} & \multicolumn{1}{c|}{\multirow{3}{1.8cm}{-\,-}} & \multicolumn{1}{c}{\multirow{3}{1.8cm}{4 \cite{ChenJJSTACS14}}} \\
\multicolumn{1}{c|}{}   &   \multicolumn{1}{c|}{}   &   \multicolumn{1}{c|}{}   & \multicolumn{1}{c|}{}  & \multicolumn{1}{c|}{} &  \multicolumn{1}{c}{} \\
\multicolumn{1}{c|}{}   &   \multicolumn{1}{c|}{}   &   \multicolumn{1}{c|}{}   & \multicolumn{1}{c|}{}  & \multicolumn{1}{c|}{} &  \multicolumn{1}{c}{} \\
\cline{1-6}
\multicolumn{1}{c|}{\multirow{2}{1.8cm}{Multiple homogeneous processors}} & \multicolumn{1}{c|}{\multirow{2}{2.0cm}{$r_j$, $d_j$, $w_j=1$, $pmtn$, $non-mig$}} & \multicolumn{1}{c|}{\multirow{2}{4.4cm}{Three states (busy, standby and off), transition cost $L_c$ is arbitrary}} & \multicolumn{1}{c|}{\multirow{2}{1.8cm}{Online}} & \multicolumn{1}{c|}{\multirow{2}{1.8cm}{-\,-}} & \multicolumn{1}{c}{\multirow{2}{3.0cm}{$O(1)$-competitive \cite{ChenJJSTACS14}}} \\
\multicolumn{1}{c|}{}   &   \multicolumn{1}{c|}{}   &   \multicolumn{1}{c|}{}   & \multicolumn{1}{c|}{}  & \multicolumn{1}{c|}{} &  \multicolumn{1}{c}{} \\
\hline\hline
\end{tabular}
\end{table*}
%%%Editor - Please note that there are two entries in column 1 that read
%%%``Multiple homogeneous processors.'' Should one of these entries be
%%%``Multiple heterogeneous processors'' instead?

For large hosting centers, several studies recommend energy-proportional computing \cite{ChaseSOSP01, Lin11, Azar13, ChabarekINFOCOM08, Bolla11, Andrews10, ZhangICNP10, HellerNSDI10}. Here, ``proportional'' refers to the concept that the size of the data center should be proportional to its workload demand \cite{Shuja12}. In \cite{ChaseSOSP01}, the authors proposed energy-conscious provisioning to concentrate the request load on a minimal active set of servers appropriate to the current aggregate load level. The excess servers would be transitioned into a power-saving state to reduce the energy cost during periods of light load. These authors reported that their system could reduce server energy usage by 29\% or more for a typical Web workload. Lin \emph{et al.} \cite{Lin11} investigated the energy-saving problem by dynamically ``right-sizing'' the target data center by turning off servers during lighter periods in both offline and online cases. They modeled the cost of toggling a server back and forth between the active and power-saving modes as a switching cost. The authors reported a 50\% energy savings for a peak-to-mean ratio (PMR) of 5. Azar \emph{et al.} \cite{Azar13} considered turning off machines to save on computation costs. The authors accounted for the delay cost as the duration required to turn the machines on or off. They proposed a bi-objective algorithm to address the problem of minimizing both the maximum task delay and the total cost.

On the network-device level, power-down mechanisms redirect network traffic to aggregate traffic on only a few network devices and links while allowing idle devices to sleep for energy conservation \cite{GuptaSIGCOMM03}. In \cite{ChabarekINFOCOM08}, the authors investigated power-aware network design and routing to dynamically power on/off line cards and chassis. They modeled the minimization of network-wide power consumption as a mixed-integer optimization problem and solved it using a branch-and-cut solver that incorporated several practical considerations (such as the network structure). They demonstrated that their results were within 11\% of the optimal result. Bolla \emph{et al.} \cite{Bolla11} used energy-aware traffic engineering on backbone networks to manage the standby and wakeup behavior of network devices with respect to the IP layer. The basic concept driving their work was to periodically reconfigure the network nodes and links to accommodate incoming traffic volumes and operational constraints (e.g., quality of service and reconvergence times). The authors noted that the energy savings depended on the level of traffic load; specifically, the energy savings was approximately 40\% when the traffic load was less than 50\%. Andrews \emph{et al.} \cite{Andrews10} studied routing and scheduling in a power-down model and captured the trade-off between energy savings and provisioning performance guarantees such as end-to-end delays. The authors proposed scheduling algorithms for line and arbitrary topologies. In the case of periodic scheduling, the schedule determined the active period per element within each frame and prioritized the packets within each active period. Zhang \emph{et al.} \cite{ZhangICNP10} proposed GreenTE, an intra-domain energy-aware traffic engineering mechanism that maximizes the number of links that can be placed in sleep mode while satisfying performance constraints (e.g., link utilization and packet delay). The authors reported an energy savings of 27-42\% for a maximum link utilization below 50\%. ElasticTree \cite{HellerNSDI10} dynamically adjusts the set of active network elements to achieve a trade-off between energy efficiency and performance. The authors used methods including a formal model, a greedy bin-packer, a topology-aware heuristic and prediction to determine which subset of links and switches to use. The authors reported a network energy savings of 50\%.
\subsection{Hybrid Technology}
Various studies of hybrid technology have also been reported that have explored both speed scaling and power-down mechanisms to fully exploit the potential for energy savings. In a standard speed scaling setting, there is a tendency to use low speed levels, subject to prescribed deadline constraints. By contrast, in combination with the power-down approach, it can be beneficial to speed up the processing of tasks to generate idle intervals in which devices can be transitioned to the power-saving mode. Irani \emph{et al.} \cite{IraniTALG07} performed the first theoretical analysis of reducing energy usage by combining placing the system in a sleep state if it is idle with varying the speed at which tasks are processed. Assuming that all tasks could be preempted and resumed at no cost, the authors proposed a $2$-approximation offline algorithm and a constant-competitive-ratio online algorithm. Several subsequent studies presented algorithms that improved on these result \cite{HanTCS10, KumarArxiv13, AlbersTALG14, AntoniadisArxiv14}. In \cite{HanTCS10}, the authors considered online scheduling of a processor to manage its energy consumption using both speed scaling and a sleep state. They proposed an algorithm called SOA to improve the competitive ratio for energy minimization from $2^{2\alpha-2}\alpha^{\alpha}+2^{\alpha-1}+2$ \cite{IraniTALG07} to $\max{\{4, \alpha^{\alpha}+2\}}$, under the assumption that the speed could be arbitrarily high. They also showed that when the maximum speed of the processor was bounded, their algorithm was $4$-competitive for throughput and $(\alpha^{\alpha}+\alpha^24^{\alpha}+2)$-competitive for energy. In \cite{KumarArxiv13}, the authors presented an algorithm called SqOA to further improve the competitive ratio of SOA \cite{HanTCS10} in the unbounded-speed setting to $\max{\{4, (2-1/{\alpha})^{\alpha}2^{\alpha-1}+2\}}$. The authors of \cite{AlbersTALG14} designed superior approximation algorithms for the offline case. They proposed a $4/3$-approximation algorithm for general convex power and a $137/117$-approximation algorithm for power functions of the form $P(s)={\beta}s^{\alpha}+\gamma$. Recently, Antoniadis \emph{et al.} \cite{AntoniadisArxiv14} closed the gap in the approximation factor by presenting a fully polynomial-time approximation scheme for the deadline-based preemptive offline scheduling problem. In \cite{LamICALP09}, the authors extended the study of this strategy to minimize the sum of energy and flow on a processor with one or multiple levels of sleep states. They designed $O(1)$-competitive clairvoyant and non-clairvoyant algorithms in the unlimited-speed setting (note: the terms ``clairvoyant'' and ``non-clairvoyant'' indicate that the sizes of the tasks are either known or not known, respectively, when the tasks arrive). Under a maximum-speed constraint, they enhanced their algorithm to again achieve $O(1)$-competitive performance.

Chen \emph{et al.} \cite{ChenSIGMETRICS05} proposed a framework to allow the employment of both approaches to enhance energy savings in hosting centers. Their solutions were based on queuing models and control theory and did not require compromise in end-user Service Level Agreements (SLAs). In \cite{YaoINFOCOM12}, the authors considered reducing the power consumed in geographically distributed data centers. They proposed an approach that incorporated selecting the number of active servers and their service rates to minimize the time-averaged power cost incurred when handling delay-tolerant workloads. Liu \emph{et al.} \cite{LiuISCA14} developed SleepScale, a power management tool to manage data centers for power efficiency while fulfilling quality-of-service (QoS) agreements. They exploited the low-power states and operating frequency that are built into modern CPUs and proposed a joint optimization of speed scaling and sleep-state selection to reduce energy consumption. \cite{NedevschiNSDI08, VasiceEnergy10} and \cite{WangICDCS14} presented studies of network elements. Nedevschi \emph{et al.} \cite{NedevschiNSDI08} explored the use of hybrid technology to reduce network energy consumption by putting network components to sleep during idle periods and adapting the rate of network operation to the requested workload. They showed that the effect depended on the power profile of the network equipment and the utilization of the network itself. Using real-world traffic workloads, topologies and power constants, they reported an energy savings of 50\% for a network utilization of 10-20\%, accompanied by a very slight increase in latency. The developers of EATe \cite{VasiceEnergy10} formulated a method of online traffic management to spread the load across multiple paths while minimizing energy consumption. Their approach leveraged the rate adaptation and power-down of links and routers to configure an energy-proportional network hardware structure. Extensive ns-2 simulations demonstrated that EATe could place 21\% of the links in the sleep state, lower the link energy levels to achieve 8\% energy savings and put 16\% of the active routers to sleep. Wang \emph{et al.} \cite{WangICDCS14} studied the problem of flow scheduling and routing while meeting flow deadlines in data center networks. They combined speed scaling and power-down mechanisms to determine routing paths and schedules as well as the transmission rate for each flow.
\subsection{Challenges and Research Issues}
The studies discussed above demonstrate that energy-efficient mechanisms can assist in making data centers green and reducing energy usage. Furthermore, they reveal that there is a trade-off between performance and energy consumption. The aforementioned efforts also have demonstrated that a considerable amount of the energy consumed as a result of over-provisioning can be saved while still satisfying performance constraints. However, there are still several other issues to be addressed concerning the design of energy-efficient algorithms:
\begin{itemize}
\item Most studies that have been conducted on energy efficiency thus far have focused on ideal power models. However, future investigations should consider scenarios that more closely resemble real systems. For example, some overhead is incurred when switching speeds because the processor must stop while the voltage is changing, and frequent changes in speed can also harm a processor{'}s lifetime reliability \cite{CoskunSIGMETRICS09}. An interesting question is how to incorporate these aspects of real systems into performance studies.
\item In classical parallel scheduling problems, the number of servers is fixed, and the scheduler primarily decides which task to process on which processor and determines the speed of each processor at a given time. However, data centers operate on a different model in which servers can be activated and released on demand \cite{BuyyaHPCC08} and in which it is necessary to exploit an appropriate number of servers to process the submitted tasks. Thus, the characteristics of data centers pose new challenges and open problems, of which the key underlying requirements are scalability, uncertainty and efficiency.
\item As mentioned earlier, the latency time and power consumption incurred in the process of rebooting means that the effects of turning servers or switches off and on cannot be overlooked. Moreover, these effects are more serious at the network level because communication data can be lost after a certain threshold latency time \cite{KuroseBookAddisonWesley13}. Fortunately, data center network architectures typically provide multiple paths for communication among their servers. The question of how to efficiently manage energy consumption in data center networks is still a major concern that is worthy of investigation. The well-structured topology of data center networks can be investigated to address this issue. A good mechanism design should minimize the effects on the network throughput, latency, congestion and packet loss while incorporating the structure of the data center network architecture.
\item Current results for multiple processors (servers) mostly focus on environments that consist of homogeneous collections of identical processors (servers). However, many computer architects believe that architectures based on heterogeneous processors will dominate future architectural design \cite{KumarRakeshPACT06, BowerIEEEMicro08}. The primary advantage of heterogeneous architectures is that the design can include certain processors that are specialized for particular types of tasks, with the intent of assigning tasks to those processors that are best suited for them \cite{GuptaSODA12}. It is also natural to consider the possibility of heterogeneity in data centers, as the various servers contained in data centers often differ from each other in their power-performance relationships. At this time, few results for heterogeneous environments are extant in energy-efficient algorithm design.
\end{itemize}

We believe that addressing the above challenges and research issues will lead to significant advances in the current level of energy efficiency of data centers.
\section{Data Center Resource Management}\label{sec:Data center resource management}
Whereas the previous section discussed schemes to increase the energy efficiency of computing and communication devices, in this section, we present prominent research works that exploit virtualization and cloud computing to efficiently utilize data center resources. The concept of virtualization is in contrast to the traditional architecture, which leads to low utilization rates and, in turn, wastes resources because servers continue to run although they are not processing applications to their full potential. Through virtualization, the power of servers can be multiplexed to address many applications, and with the help of virtualization, more than one application can be assigned to one server. Thus, this technique enables significant and cost-effective gains in resource utilization and associated energy savings. A second innovation that can greatly increase utilization rates is the development and maturation of cloud computing. The shift to the cloud encourages economies of scale because servers can be run on virtual platforms, which offer higher utilization rates, and these large cloud providers are highly incentivized to optimize the utilization of data centers and reduce energy consumption to maximize productivity. In this section, we analyze methods that adopt these two approaches for virtual machine assignment, network traffic engineering, power distribution and renewable energy access with a view to providing better energy management for green data centers, and we consider the corresponding directions for future research.
\subsection{Energy-Aware Virtual Machine Assignment}
Underutilization in data centers is a major cause of excessive energy consumption. Moreover, modern servers are sufficiently powerful to use virtualization to present the illusion of many smaller virtual machines (VMs), each running independent applications \cite{BarhamSOSP03}. Therefore, optimal virtual machine management (including allocation, consolidation, and migration) is an important technique for data centers that can facilitate the improvement of resource utilization and the reduction of energy consumption. The general idea is to efficiently map VMs to servers based on resource matching and improved performance metrics. An early study \cite{NathujiSOSP07} extended virtualization solutions to support rich and effective policies for active power management, which had not been done before. The authors integrated ``hard'' and ``soft'' power states to achieve high power savings and showed that substantial benefits could be derived from the coordination of online methods for server consolidation using their proposed management techniques. In \cite{RaghavendraASPLOS08}, Raghavendra \emph{et al.} proposed a power management solution that coordinated various individual approaches, including a virtual machine controller. Their design used techniques and actuators to optimize power at multiple system levels across both hardware and software. Hermenier \emph{et al.} \cite{HermenierVEE09} proposed Entropy, a consolidation manager that performed dynamic consolidation of VMs based on constraint programming and accounted for migration overhead. The authors showed that their approach could significantly reduce the number of nodes and the migration time compared with results obtained using the FFD heuristic. They reported that the consumption of nodes per hour could be reduced by 25\%.

When exploiting the ability to consolidate several virtual machines on the same physical server, a key challenge is to explicitly account for the specified QoS in the optimization problem \cite{BobroffIM07, WoodNSDI07}. Van \emph{et al.} \cite{VanCIT09} defined the condition of SLA fulfillment and the resource management cost as a global utility function. The authors adopted the Constraint Programming approach to formulate and solve the considered problem. Kusic \emph{et al.} \cite{KusicCluster09} considered the problem of consolidating services into a smaller number of computing resources to achieve higher server utilization and energy efficiency while maintaining the desired QoS. The authors implemented a dynamic resource provisioning framework to treat this problem as a problem of sequential optimization and solve it using a lookahead control scheme. They reported 22\% power savings compared with a system without dynamic control. Several other efforts have also been directed toward the online assignment of virtual machines to fewer servers. Beloglazov \emph{et al.} \cite{BeloglazovConCom11} proposed novel adaptive heuristics for the dynamic consolidation of VMs based on an analysis of historical data related to VM resource usage. They applied a modification of the Best Fit Decreasing algorithm to perform the VM assignment, and they achieved significant reduction of energy consumption while ensuring a high level of adherence to the SLA. In \cite{WangINFOCOM11}, the authors considered the problem of consolidating virtual machines when the network bandwidth demands of the VMs are dynamic. They converted the problem into a Stochastic Bin Packing problem and proposed an online packing algorithm. They reported that the performance ratio was within $(1+\epsilon)(\sqrt{2}+1)$ for any $\epsilon>0$ and that the number of required servers was reduced by 30\% compared with the results of the HARMONIC algorithm \cite{LeeCCJACM85}. In \cite{StolyarSIGMETRICS13}, policies that minimized the total number of occupied servers in the system were studied. The authors presented Greedy with sublinear Safety Stocks (GSS) and Modified GSS (GSS-M) policies. Generally speaking, these GSS and GSS-M policies scheduled incoming virtual machines in a manner that greedily minimized the total number of occupied physical servers. The authors demonstrated the asymptotic optimality and convergence rate of their policies. However, such beneficial consolidation also introduces an SLA violation risk. In \cite{JinGlobeCom13}, the authors presented a risk management framework to address the problem of consolidating virtual machines for energy efficiency in data centers. They suggested a two-step approach to virtual machine scheduling: 1) they solved the problem of resource allocation for virtual machines using a stochastic program, and 2) they proposed an algorithm for dynamic virtual machine consolidation at run time. Using real-world workloads, they showed that SLA violations were reduced by a factor of four, from 25\% to 2-5\%, compared with other virtual machine consolidation algorithms that do not consider the SLA violation risk.

Table~\ref{tb:Comparison of the proposals regarding virtual machine assignment} summarizes the various approaches to energy-aware virtual machine assignment and presents a comparison of the common features identified among the different proposals mentioned above. The features selected for comparison are as follows: 1) the resources considered, 2) the techniques used in the proposals, 3) the traces used for the experiments, 4) the performance metrics considered, and 5) the results of the proposals.
\renewcommand{\arraystretch}{1.3}
\begin{table*}[htbp]
\centering
\scriptsize
\caption{\label{tb:Comparison of the proposals regarding virtual machine assignment} Comparison of the proposals regarding virtual machine assignment}
\begin{tabular}{p{2.3cm}|p{2.2cm}|p{3.0cm}|p{2.9cm}|p{2.5cm}|p{2.1cm}}
%{c|p{2.5cm}|c|p{2.5cm}|c|p{3.5cm}|c|p{3.5cm}|c|p{3.5cm}|c|p{3.5cm}}
\hline\hline
Proposal & VM Resources & Techniques & Traces & Performance & Result \\\hline
\multicolumn{1}{c|}{\multirow{3}{2.3cm}{VirtualPower \cite{NathujiSOSP07}}} & 
\multicolumn{1}{c|}{\multirow{3}{2.2cm}{Cores on chip}} & 
\multicolumn{1}{c|}{\multirow{3}{3.0cm}{VM consolidation, DVFS, resource (time slice) scheduling, server power-down}} & 
\multicolumn{1}{c|}{\multirow{3}{2.9cm}{RUBiS tiered web service, transactional load}} & 
\multicolumn{1}{c|}{\multirow{3}{2.5cm}{Processing rate, throughput}} & 
\multicolumn{1}{c}{\multirow{3}{2.1cm}{34\% reduction in power}} \\
\multicolumn{1}{c|}{}   &   \multicolumn{1}{c|}{}   &   \multicolumn{1}{c|}{}   & \multicolumn{1}{c|}{}  & \multicolumn{1}{c|}{} &  \multicolumn{1}{c}{} \\
\multicolumn{1}{c|}{}   &   \multicolumn{1}{c|}{}   &   \multicolumn{1}{c|}{}   & \multicolumn{1}{c|}{}  & \multicolumn{1}{c|}{} &  \multicolumn{1}{c}{} \\

\hline
\multicolumn{1}{c|}{\multirow{2}{2.3cm}{No ``Power" Struggles \cite{RaghavendraASPLOS08}}} & 
\multicolumn{1}{c|}{\multirow{2}{2.2cm}{CPU}} & 
\multicolumn{1}{c|}{\multirow{2}{3.0cm}{VM consolidation, DVFS, server power-down}} & 
\multicolumn{1}{c|}{\multirow{2}{2.9cm}{Trace from enterprise deployment}} & 
\multicolumn{1}{c|}{\multirow{2}{2.5cm}{Throughput, power budget}} & 
\multicolumn{1}{c}{\multirow{2}{2.1cm}{64\% reduction in power}} \\
\multicolumn{1}{c|}{}   &   \multicolumn{1}{c|}{}   &   \multicolumn{1}{c|}{}   & \multicolumn{1}{c|}{}  & \multicolumn{1}{c|}{} &  \multicolumn{1}{c}{} \\

\hline
\multicolumn{1}{c|}{\multirow{2}{2.3cm}{Entropy \cite{HermenierVEE09}}} & 
\multicolumn{1}{c|}{\multirow{2}{2.2cm}{Processing unit, memory}} & 
\multicolumn{1}{c|}{\multirow{2}{3.0cm}{VM consolidation}} & 
\multicolumn{1}{c|}{\multirow{2}{2.9cm}{NASGrid benchmark suite}} & 
\multicolumn{1}{c|}{\multirow{2}{2.5cm}{Migration time}} & 
\multicolumn{1}{c}{\multirow{2}{2.1cm}{25\% reduction in nodes per hour}} \\
\multicolumn{1}{c|}{}   &   \multicolumn{1}{c|}{}   &   \multicolumn{1}{c|}{}   & \multicolumn{1}{c|}{}  & \multicolumn{1}{c|}{} &  \multicolumn{1}{c}{} \\

\hline
\multicolumn{1}{c|}{\multirow{2}{2.3cm}{Resource management \cite{VanCIT09}}} & 
\multicolumn{1}{c|}{\multirow{2}{2.2cm}{CPU, memory}} & 
\multicolumn{1}{c|}{\multirow{2}{3.0cm}{VM provisioning, VM packing}} & 
\multicolumn{1}{c|}{\multirow{2}{2.9cm}{Multiplayer online game, web application}} & 
\multicolumn{1}{c|}{\multirow{2}{2.5cm}{Response time, throughput}} & 
\multicolumn{1}{c}{\multirow{2}{2.1cm}{Trade-off between SLA and energy}} \\
\multicolumn{1}{c|}{}   &   \multicolumn{1}{c|}{}   &   \multicolumn{1}{c|}{}   & \multicolumn{1}{c|}{}  & \multicolumn{1}{c|}{} &  \multicolumn{1}{c}{} \\

\hline
\multicolumn{1}{c|}{\multirow{2}{2.3cm}{Power {\&} SLA management \cite{KusicCluster09}}} & 
\multicolumn{1}{c|}{\multirow{2}{2.2cm}{CPU}} & 
\multicolumn{1}{c|}{\multirow{2}{3.0cm}{VM consolidation, DVFS, server power-down}} & 
\multicolumn{1}{c|}{\multirow{2}{2.9cm}{Web workload, Trade6 enterprise application}} & 
\multicolumn{1}{c|}{\multirow{2}{2.5cm}{Response time}} & 
\multicolumn{1}{c}{\multirow{2}{2.1cm}{22\% reduction in power}} \\
\multicolumn{1}{c|}{}   &   \multicolumn{1}{c|}{}   &   \multicolumn{1}{c|}{}   & \multicolumn{1}{c|}{}  & \multicolumn{1}{c|}{} &  \multicolumn{1}{c}{} \\

\hline
\multicolumn{1}{c|}{\multirow{2}{2.3cm}{Consolidation of VM \cite{BeloglazovConCom11}}} & 
\multicolumn{1}{c|}{\multirow{2}{2.2cm}{CPU}} & 
\multicolumn{1}{c|}{\multirow{2}{3.0cm}{Dynamic consolidation of VMs}} & 
\multicolumn{1}{c|}{\multirow{2}{2.9cm}{Trace from PlanetLab}} & 
\multicolumn{1}{c|}{\multirow{2}{2.5cm}{Capacity violation, VM migration}} & 
\multicolumn{1}{c}{\multirow{2}{2.1cm}{Significant reduction in energy}} \\
\multicolumn{1}{c|}{}   &   \multicolumn{1}{c|}{}   &   \multicolumn{1}{c|}{}   & \multicolumn{1}{c|}{}  & \multicolumn{1}{c|}{} &  \multicolumn{1}{c}{} \\

\hline
\multicolumn{1}{c|}{\multirow{2}{2.3cm}{Consolidation VMs \cite{WangINFOCOM11}}} & 
\multicolumn{1}{c|}{\multirow{2}{2.2cm}{Bandwidth}} & 
\multicolumn{1}{c|}{\multirow{2}{3.0cm}{VM consolidation}} & 
\multicolumn{1}{c|}{\multirow{2}{2.9cm}{Trace from IBM Global service and cluster}} & 
\multicolumn{1}{c|}{\multirow{2}{2.5cm}{Capacity violation}} & 
\multicolumn{1}{c}{\multirow{2}{2.1cm}{30\% reduction in required number of servers}} \\
\multicolumn{1}{c|}{}   &   \multicolumn{1}{c|}{}   &   \multicolumn{1}{c|}{}   & \multicolumn{1}{c|}{}  & \multicolumn{1}{c|}{} &  \multicolumn{1}{c}{} \\

\hline
\multicolumn{1}{c|}{\multirow{2}{2.3cm}{Minimize occupied servers \cite{StolyarSIGMETRICS13}}} & 
\multicolumn{1}{c|}{\multirow{2}{2.2cm}{CPU, memory}} & 
\multicolumn{1}{c|}{\multirow{2}{3.0cm}{VM placement with packing constraints}} & 
\multicolumn{1}{c|}{\multirow{2}{2.9cm}{-\,-}} & 
\multicolumn{1}{c|}{\multirow{2}{2.5cm}{-\,-}} & 
\multicolumn{1}{c}{\multirow{2}{2.1cm}{Reduced number of servers}} \\
\multicolumn{1}{c|}{}   &   \multicolumn{1}{c|}{}   &   \multicolumn{1}{c|}{}   & \multicolumn{1}{c|}{}  & \multicolumn{1}{c|}{} &  \multicolumn{1}{c}{} \\

\hline
\multicolumn{1}{c|}{\multirow{3}{2.3cm}{Risk management for VMs \cite{JinGlobeCom13}}} & 
\multicolumn{1}{c|}{\multirow{3}{2.2cm}{CPU, memory}} & 
\multicolumn{1}{c|}{\multirow{3}{3.0cm}{Resource allocation, VM consolidation}} & 
\multicolumn{1}{c|}{\multirow{3}{2.9cm}{Trace from CoMon project}} & 
\multicolumn{1}{c|}{\multirow{3}{2.5cm}{Capacity violation}} & 
\multicolumn{1}{c}{\multirow{3}{2.1cm}{Factor of 4 reduction in SLA violations}} \\
\multicolumn{1}{c|}{}   &   \multicolumn{1}{c|}{}   &   \multicolumn{1}{c|}{}   & \multicolumn{1}{c|}{}  & \multicolumn{1}{c|}{} &  \multicolumn{1}{c}{} \\
\multicolumn{1}{c|}{}   &   \multicolumn{1}{c|}{}   &   \multicolumn{1}{c|}{}   & \multicolumn{1}{c|}{}  & \multicolumn{1}{c|}{} &  \multicolumn{1}{c}{} \\
\hline\hline
\end{tabular}
\end{table*}
\subsection{Data Center Network Traffic Engineering}
Traffic engineering is a highly effective approach to making data centers green based on different traffic patterns and network architectures \cite{MahadevanMagazine11}. Several solutions have been proposed for achieving energy proportionality by applying traffic aggregation and virtual machine assignment techniques. In \cite{McGeerGlobecom10}, the authors demonstrated that the minimization of network power consumption for any general data center topology is NP-complete. They proposed a centralized network power controller program that gathered traffic data and server statistics from all switches and servers and used this information to perform traffic aggregation and virtual machine assignment and migration in the target data center. They demonstrated the optimal placement for the virtual machines and the bandwidth savings through trace evaluation. In \cite{VasicCoNEXT11}, the authors developed REsPoNse, a framework based on identifying and using energy-critical paths to redirect traffic to allow large portions of the network to enter a low-power state. They demonstrated that energy savings were achieved without frequent recomputation and deployment of new routing tables, and they showed that REsPoNse had marginal impact on the application-level throughput and latency. Their results yielded energy savings of approximately 30\% and 42\%. Jiang \emph{et al.} \cite{JiangINFOCOM12} considered a joint design incorporating virtual machine assignment and routing for data centers. They considered the optimization of both energy cost and network congestion, then proposed an offline algorithm based on a Markov chain model and an incremental online solution for a dynamic environment. The authors reported performance analysis results for various topologies, including clique, fat-tree, HyperX and BCube topologies, under a spectrum of elephant and mice workloads. Given that the bandwidth demands of different flows do not peak at exactly the same time, the authors \cite{WangINFOCOM12} proposed a correlation-aware power optimization scheme that dynamically consolidated traffic flows onto a small set of switches and links and then shut down unused network devices. They designed a heuristic algorithm to determine the consolidation and rate configuration, which was fundamentally based on a bin-packing algorithm. They reported an energy savings of 46\% in network energy for a data center network with only a negligible increase in delay. To exploit the well-structured topology of data centers, Zhang \emph{et al.} \cite{ZhangICC12} adopted a hierarchical perspective to study the task of optimizing power usage in data center networks while guaranteeing connectivity and maximum link utilization. They divided their discussion into two levels: 1) the core level, for the determination of the core switches that must be powered to carry outgoing traffic and the aggregation switches to serve the out-pod traffic in each pod, and 2) the pod level, for the determination of the aggregation switches that must be active to carry intra-pod traffic. They designed a hierarchical energy optimization algorithm and evaluated it on various traffic patterns, including large traffic flow only, small traffic flow only, and random traffic patterns. In \cite{XuNetworks13}, the authors adopted the routing perspective to consider reducing the power consumption of high-density data center networks while satisfying network performance requirements. They proposed a routing algorithm that used as little network power as possible to provide routing services, and they reported a 39\% energy savings for a link load of 20\% and a reliability requirement of 2 (i.e., the threshold for the ratio of the network throughput that the data center operator could tolerate after switch and link elimination to the basic throughput achievable using all switches was 2). In \cite{WangJSAC14}, the authors proposed a new framework to embrace the new opportunities offered by combining certain unique features of data centers with traffic engineering. They solved the problem in two steps: first, by allocating virtual machines to servers to reduce the amount of traffic and to generate favorable conditions for traffic engineering, and then, by reducing the number of active switches and balancing traffic flows to reduce energy consumption. The experimental results indicated an energy savings of up to 50\%. In addition to considering the network level, the authors of \cite{JinIPDPS13} considered a joint host-network energy optimization scheme involving the simultaneous optimization of virtual machine assignment and network flow routing. They modeled the joint optimization problem as an integer linear program. Because this problem is NP-complete, they proposed a series of techniques to address the associated challenges: 1) they converted the problem to one that considered a single optimization solution, 2) they proposed a parallelization approach for rapid completion, and 3) they quickly identified efficient routing paths for the flows. The authors reported that their optimization method reduced energy consumption by 40\%  for a system load of 30\%. By utilizing correlation analysis in the joint power optimization of a data center network and servers, the authors of \cite{ZhengINFOCOM14} proposed PowerNetS to achieve improved energy savings. Their approach featured a well-designed heuristic algorithm to incrementally perform virtual machine and traffic consolidation with lower virtual machine migration overheads, thereby optimizing power consumption while guaranteeing the fulfillment of network constraint. They reported that their method could achieve 51.6\% energy savings for a data center.

In Table~\ref{tb:Comparison of the proposals regarding network traffic engineering}, we summarize and compare the proposals mentioned above. The  features presented for comparison are 1) the traffic patterns, 2) the network architectures, 3) the techniques used in the proposals, 4) the considered QoS specifications, and 5) the results of the proposals.
\renewcommand{\arraystretch}{1.3}
\begin{table*}[htbp]
\centering
\scriptsize
\caption{\label{tb:Comparison of the proposals regarding network traffic engineering} Comparison of the proposals regarding network traffic engineering}
\begin{tabular}{p{2.1cm}|p{2.2cm}|p{2.8cm}|p{2.5cm}|p{3.0cm}|p{2.2cm}}
\hline\hline
Proposal & Traffic Patterns & Network Architectures & Techniques & QoS & Result \\\hline
\multicolumn{1}{c|}{\multirow{2}{2.1cm}{Network power controller \cite{McGeerGlobecom10}}} & 
\multicolumn{1}{c|}{\multirow{2}{2.2cm}{E-commerce trace}} & 
\multicolumn{1}{c|}{\multirow{2}{2.8cm}{Fat tree}} & 
\multicolumn{1}{c|}{\multirow{2}{2.8cm}{Traffic aggregation, VM placement}} & 
\multicolumn{1}{c|}{\multirow{2}{3.1cm}{Bandwidth, connectivity guarantees}} & 
\multicolumn{1}{c}{\multirow{2}{2.2cm}{55\% network power reduction}} \\
\multicolumn{1}{c|}{}   &   \multicolumn{1}{c|}{}   &   \multicolumn{1}{c|}{}   & \multicolumn{1}{c|}{}  & \multicolumn{1}{c|}{} &  \multicolumn{1}{c}{} \\

\hline
\multicolumn{1}{c|}{\multirow{2}{2.1cm}{REsPoNse \cite{VasicCoNEXT11}}} & 
\multicolumn{1}{c|}{\multirow{2}{2.2cm}{Media stream, web workload}} & 
\multicolumn{1}{c|}{\multirow{2}{2.8cm}{Fat tree}} & 
\multicolumn{1}{c|}{\multirow{2}{2.8cm}{Energy critical-path detection}} & 
\multicolumn{1}{c|}{\multirow{2}{3.1cm}{Throughput, latency}} & 
\multicolumn{1}{c}{\multirow{2}{2.2cm}{40\% network energy savings}} \\
\multicolumn{1}{c|}{}   &   \multicolumn{1}{c|}{}   &   \multicolumn{1}{c|}{}   & \multicolumn{1}{c|}{}  & \multicolumn{1}{c|}{} &  \multicolumn{1}{c}{} \\

\hline
\multicolumn{1}{c|}{\multirow{2}{2.1cm}{Joint controller \cite{JiangINFOCOM12}}} & 
\multicolumn{1}{c|}{\multirow{2}{2.2cm}{Parallel workload}} & 
\multicolumn{1}{c|}{\multirow{2}{2.8cm}{Fat tree, clique, HyperX, BCube}} & 
\multicolumn{1}{c|}{\multirow{2}{2.8cm}{Route selection, VM placement}} & 
\multicolumn{1}{c|}{\multirow{2}{3.1cm}{Congestion}} & 
\multicolumn{1}{c}{\multirow{2}{2.2cm}{Data center cost improvement}} \\
\multicolumn{1}{c|}{}   &   \multicolumn{1}{c|}{}   &   \multicolumn{1}{c|}{}   & \multicolumn{1}{c|}{}  & \multicolumn{1}{c|}{} &  \multicolumn{1}{c}{} \\

\hline
\multicolumn{1}{c|}{\multirow{2}{2.1cm}{CARPO \cite{WangINFOCOM12}}} & 
\multicolumn{1}{c|}{\multirow{2}{2.2cm}{Wikipedia trace}} & 
\multicolumn{1}{c|}{\multirow{2}{2.8cm}{Fat tree}} & 
\multicolumn{1}{c|}{\multirow{2}{2.8cm}{Traffic consolidation, link-rate adaptation}} & 
\multicolumn{1}{c|}{\multirow{2}{3.1cm}{Packet delay}} & 
\multicolumn{1}{c}{\multirow{2}{2.2cm}{46\% network energy savings}} \\
\multicolumn{1}{c|}{}   &   \multicolumn{1}{c|}{}   &   \multicolumn{1}{c|}{}   & \multicolumn{1}{c|}{}  & \multicolumn{1}{c|}{} &  \multicolumn{1}{c}{} \\

\hline
\multicolumn{1}{c|}{\multirow{2}{2.1cm}{HERO \cite{ZhangICC12}}} & 
\multicolumn{1}{c|}{\multirow{2}{2.2cm}{Large, small and random traffic}} & 
\multicolumn{1}{c|}{\multirow{2}{2.8cm}{Fat tree}} & 
\multicolumn{1}{c|}{\multirow{2}{2.8cm}{Switching off network switches}} & 
\multicolumn{1}{c|}{\multirow{2}{3.1cm}{Connectivity, maximum link utilization}} & 
\multicolumn{1}{c}{\multirow{2}{2.2cm}{network power savings}} \\
\multicolumn{1}{c|}{}   &   \multicolumn{1}{c|}{}   &   \multicolumn{1}{c|}{}   & \multicolumn{1}{c|}{}  & \multicolumn{1}{c|}{} &  \multicolumn{1}{c}{} \\

\hline
\multicolumn{1}{c|}{\multirow{3}{2.1cm}{PRP \cite{XuNetworks13}}} & 
\multicolumn{1}{c|}{\multirow{3}{2.2cm}{one-to-one, one-to-many, all-to-all}} & 
\multicolumn{1}{c|}{\multirow{3}{2.8cm}{Fat tree, BCube}} & 
\multicolumn{1}{c|}{\multirow{3}{2.8cm}{Switch and link elimination}} & 
\multicolumn{1}{c|}{\multirow{3}{3.1cm}{Throughput}} & 
\multicolumn{1}{c}{\multirow{3}{2.2cm}{39\% network energy savings}} \\
\multicolumn{1}{c|}{}   &   \multicolumn{1}{c|}{}   &   \multicolumn{1}{c|}{}   & \multicolumn{1}{c|}{}  & \multicolumn{1}{c|}{} &  \multicolumn{1}{c}{} \\
\multicolumn{1}{c|}{}   &   \multicolumn{1}{c|}{}   &   \multicolumn{1}{c|}{}   & \multicolumn{1}{c|}{}  & \multicolumn{1}{c|}{} &  \multicolumn{1}{c}{} \\

\hline
\multicolumn{1}{c|}{\multirow{3}{2.1cm}{GreenDCN \cite{WangJSAC14}}} & 
\multicolumn{1}{c|}{\multirow{3}{2.2cm}{Communication-intensive}} & 
\multicolumn{1}{c|}{\multirow{3}{2.8cm}{Fat tree}} & 
\multicolumn{1}{c|}{\multirow{3}{2.8cm}{VM assignment, switch and link engineering}} & 
\multicolumn{1}{c|}{\multirow{3}{3.1cm}{Connectivity}} & 
\multicolumn{1}{c}{\multirow{3}{2.2cm}{50\% network energy savings}} \\
\multicolumn{1}{c|}{}   &   \multicolumn{1}{c|}{}   &   \multicolumn{1}{c|}{}   & \multicolumn{1}{c|}{}  & \multicolumn{1}{c|}{} &  \multicolumn{1}{c}{} \\
\multicolumn{1}{c|}{}   &   \multicolumn{1}{c|}{}   &   \multicolumn{1}{c|}{}   & \multicolumn{1}{c|}{}  & \multicolumn{1}{c|}{} &  \multicolumn{1}{c}{} \\

\hline
\multicolumn{1}{c|}{\multirow{2}{2.1cm}{Joint optimization \cite{JinIPDPS13}}} & 
\multicolumn{1}{c|}{\multirow{2}{2.2cm}{Iperf UDP}} & 
\multicolumn{1}{c|}{\multirow{2}{2.8cm}{Fat tree}} & 
\multicolumn{1}{c|}{\multirow{2}{2.8cm}{VM placement, flow routing}} & 
\multicolumn{1}{c|}{\multirow{2}{3.1cm}{-\,-}} & 
\multicolumn{1}{c}{\multirow{2}{2.2cm}{40\% energy savings}} \\
\multicolumn{1}{c|}{}   &   \multicolumn{1}{c|}{}   &   \multicolumn{1}{c|}{}   & \multicolumn{1}{c|}{}  & \multicolumn{1}{c|}{} &  \multicolumn{1}{c}{} \\

\hline
\multicolumn{1}{c|}{\multirow{3}{2.1cm}{PowerNetS \cite{ZhengINFOCOM14}}} & 
\multicolumn{1}{c|}{\multirow{3}{2.2cm}{Wikipedia, Yahoo! and IBM traces}} & 
\multicolumn{1}{c|}{\multirow{3}{2.8cm}{Fat tree}} & 
\multicolumn{1}{c|}{\multirow{3}{2.8cm}{VM and flow consolidation}} & 
\multicolumn{1}{c|}{\multirow{3}{3.1cm}{Packet delay}} & 
\multicolumn{1}{c}{\multirow{3}{2.2cm}{51.6\% energy savings}} \\
\multicolumn{1}{c|}{}   &   \multicolumn{1}{c|}{}   &   \multicolumn{1}{c|}{}   & \multicolumn{1}{c|}{}  & \multicolumn{1}{c|}{} &  \multicolumn{1}{c}{} \\
\multicolumn{1}{c|}{}   &   \multicolumn{1}{c|}{}   &   \multicolumn{1}{c|}{}   & \multicolumn{1}{c|}{}  & \multicolumn{1}{c|}{} &  \multicolumn{1}{c}{} \\
\hline\hline
\end{tabular}
\end{table*}
\subsection{Power Distribution}
Whereas virtual machine consolidation allows the total number of active physical servers to be reduced, power distribution approaches have also been proposed to reduce the high energy costs of idle systems. Driven by increasing power density, the increasing number of IT devices, and the capability of on-demand addition and removal of IT devices, there have been dramatic changes in how power is utilized in data centers \cite{RasmussenAPC09}. In general, power distribution is over-provisioned in data centers to accommodate peaks and to allow for future expansion. However, based on the assumption that simultaneous peak draw across all equipment will occur only rarely, power over-subscription is intentionally exploited to improve power utilization in modern data centers \cite{Barroso13, FanISCA07, PelleyASPLOS10, LimUSENIXATC11}. Fan \emph{et al.} \cite{FanISCA07} were the first to study power usage at the scale of data center workloads, and they reported the use of power modeling for power provisioning. Their work yielded several key findings: 1) The gap between the maximum power actually used by large groups of servers and their aggregate theoretical peak usage can be as large as 40\% in data centers, which suggests a significant opportunity to host additional servers under the same power budget; 2) power capping is more useful as a safety mechanism to prevent overload situations; 3) large groups of servers are operating near peak power levels in certain time intervals, which suggests that power gaps and power management techniques can be more easily exploited at the data center level than at the rack level; and 4) when large groups of servers are considered, frequency scaling has the potential to be moderately effective at reducing peak power consumption. Their evaluation results indicated energy savings of 35-40\%. Meisner \emph{et al.} \cite{MeisnerASPLOS09} proposed PowerNap, an energy-conservation approach that supported an entire system in transitioning rapidly between a high-performance active state and a near-zero-power idle state in response to instantaneous variations in load. Based on the PowerNap concept, the authors developed requirements and mechanisms to eliminate idle power waste in enterprise blade servers. The authors also introduced another power provisioning approach named RAILS to be applied when PowerNap was operating in the low-efficiency regions of current blade center power supplies with the intent of providing high conversion efficiency throughout the entire range. They reported that PowerNap and RAILS could reduce the average server power consumption by 74\%. To overcome local utilization spikes that could force the throttling of server performance to enforce safe power budgets, Pelley \emph{et al.} \cite{PelleyASPLOS10} developed mechanisms to better utilize the installed power infrastructure. They proposed Power Routing to centrally schedule servers dynamically across redundant power feeds, and they proved that the problem of assigning servers to power distribution units (PDUs) was NP-complete. The authors first optimally solved the relaxation problem through linear programming, assuming the servers to be assigned fractionally across feeds, and then constructed an approximate solution to the original problem. Using traces from production systems, they demonstrated that capital costs could be reduced by 32\% without performance degradation. For the case in which servers are shared by virtual machines belonging to different applications, a power budgeting solution named VPS was presented by \cite{LimUSENIXATC11} to guarantee the quality of service. VPS dynamically shifted power among various distributed components as workloads and power availability varied to efficiently utilize the power budget. The authors combined hardware-based and software-based power control knobs to optimize the performance. They used traces to evaluate the following performance metrics: data-center-level and application-level power budgeting errors, application performance differentiation, and performance achieved within the power budget. In \cite{IsciISCA13}, the authors explored the feasibility of low-latency power states for enterprise server systems and proposed an end-to-end virtualization power management solution. They implemented the capability of a low-power and low-latency state for virtualization servers and demonstrated this technology could overcome the traditional barriers (e.g., incurred latencies) to achieve substantial energy savings via dynamic consolidation. Based on real-system evaluations, they reported that data center power efficiency could be improved by 30\%.

In addition to power capping, several studies have introduced Uninterruptible Power Supplies (UPSs) to store energy and reduce power budget violations \cite{GovindanISCA11, KontorinisISCA12, GovindanASPLOS12}. In \cite{GovindanISCA11}, the authors pioneered the leveraging of a UPS in a data center as an energy buffer (eBuff) to address the peak power draw problem. They used the energy stored in UPS batteries to provide energy during peak demand, resulting in a reduction in capital expenses and operating expenses without performance degradation. They reported that eBuff yielded a 15-45\% peak reduction, which corresponded to a 6-18\% savings in operation expenses. Because eBuff was centralized and limited by the capacity of the UPS battery, Kontorinis \emph{et al.} \cite{KontorinisISCA12} presented an architecture for distributed per-server UPSs that stored energy during low-activity periods and used this energy during power spikes. They proposed details regarding the sizing as well as technological alternatives and approaches that managed battery charging and discharging behavior while addressing reliability and availability concerns. They demonstrated that their policies prolonged the duration of UPS batteries' usage and reduced their total cost of ownership per server by 6.3\%. In \cite{GovindanASPLOS12}, the authors studied the possibility of leveraging stored energy to address power emergencies. Their work demonstrated that mechanisms for power capping had performance-degrading ramifications, and they proposed an offline theoretical framework and several online heuristics to temporarily augment utility supplies during emergencies. They reported that battery-based solutions 1) could cope with emergencies of short duration, 2) supplemented existing reaction strategies to enhance efficacy for longer emergencies, and 3) provided feasible options when other strategies (power states, migration, etc.) did not suffice. In the case of aggressive over-subscription, whereas the aforementioned work suggested power capping and stored energy to reduce the risk of power peaks, the authors of \cite{WangISCA13} suggested that the power hierarchy should be treated as a data center resource of comparable importance to that of other computing resources. To this end, the authors proposed vPower, a software system to virtualize power distribution. vPower combined computing parameters (such as DVFS, consolidation, etc.) and energy storage devices to establish a virtual power hierarchy for each application. The primary advantages of vPower were that it 1) allowed applications to specify their power needs, 2) performed admission control and assignment, 3) dynamically monitored power usage, and 4) enforced the allocation of power for fairness and system efficiency. The results of a power hierarchy prototype indicated the achieved improvement in system utilization was over 50\% compared with the conservative scheduling approach based on peak load requirements.  
\subsection{Renewable Energy Access}
Expanding the use of renewable energy is another route toward green data centers. Some data center owners have already investigated and accepted alternative energies, such as solar, geothermal and wind energy \cite{StoneTimes07, GuptaTimes10}. However, utilizing these green energies is challenging because of their restricted, intermittent and unstable nature. Therefore, the main question that must be addressed is how to exploit renewable energy and overcome the associated hindrances by exploiting the unique features of data centers. Sharma \emph{et al.} \cite{SharmaASPLOS11} developed Blink to address intermittent power constraints. Based on workload characteristics and energy constraints, Blink determined the power state of the servers and minimized performance degradation. Because the power varied and the energy stored in the batteries was for short-term use, Blink's policy had potential benefits for distributed applications. The authors constructed a cluster of 10 low-power motherboards powered by an array of micro wind turbines and solar panels to evaluate the performance under intermittent power constraints. They demonstrated that an asymmetric load-proportional policy could increase fairness without significantly sacrificing the cache's hit rate to achieve a blinking version of memcached. Considering the location restrictions of renewable energy, Akoush \emph{et al.} \cite{AkoushHotOS11} proposed a computation architecture to avoid the necessity of expensive infrastructure investment. Their design consisted of three relevant components: 1) data centers collocated with renewable energy sources, 2) data centers linked together over a dedicated communication network, and 3) a software framework that supported the seamless execution and migration of virtual machines depending on power availability. Although they demonstrated the viability of the design in a case study, the authors also noted the technical challenges and inherent limitations of their approach, such as the efficiency of virtual machine migration between data centers and suitable workload concerns. In \cite{LiISCA13}, the authors leveraged tunable energy, such as that provided by fuel cells and gas turbines, to propose a power management approach that enabled data center operation with high performance and low overhead. Unlike previous work, in which the workload had been forced to track the variable power budget, their approach explored the possibility of achieving a renewable energy supply to follow the data center power demand. To overcome issues of performance degradation, they proposed two adaptive load tuning schemes, named DGR Boost and UPS Boost, based on mature computer tuning approaches (such as DVFS and UPS batteries). Evaluations performed on real-world data center traces and industry data on distributed generation systems indicated that their technique achieved a 37\% improvement in performance over existing supply-tracking-based designs and saved $100$ metric tons of carbon emissions annually, given that the power consumption of the data center was $10 MW$.

If powered only by variable sources of green energy, a data center will experience unpredictable performance for applications. The most reasonable approach is to incorporate renewable energy into the electrical grid. In \cite{GoiriSC11} and \cite{GoiriEuroSys12}, the authors considered a data center powered by a solar array and the electrical grid. Two scheduler frameworks, named GreenSlot and GreenHadoop, were proposed for parallel batch jobs and data-processing jobs, respectively. Both schedulers used historical data and weather forecasts to predict the amount of solar energy that would be available in the future, with the intent of maximizing the use of green energy. Because of the inherent natural variability in the availability of solar energy, grid energy was used to avoid deadline violations. The results obtained using a real-world trace demonstrated that GreenSlot could increase green energy consumption by up to 117\% compared with a conventional backfilling scheduler, whereas GreenHadoop could increase green energy consumption by up to 31\% compared with the Hadoop algorithm. However, the inefficient and redundant load-matching activities of these algorithms can incur performance losses. iSwith \cite{LiISCA12} explored the design trade-offs between energy utilization and load tuning (e.g., DVFS and CPU power states) for a data center partially powered by intermittent renewable energy. iSwith combined optimization of the supply-side fluctuation with minimization of the load-fluctuation-induced overhead to mitigate performance loss. This system could reduce job waiting time by 80\% and also mitigate peak network traffic by 95\% and average network traffic by 75\% while still maintaining 96\% renewable energy utilization. 

To avoid transmitting renewable energy over long distances, which leads to significant attenuation losses \cite{AkoushHotOS11}, several proposed solutions have suggest dispatching traffic to multiple data centers in different geographical locations to use the renewable energy available nearby \cite{LiuSIGMETRICS11, GaoINFOCOM13, GaoSIGCOMM12, ZhangMiddleware11}. In \cite{LiuSIGMETRICS11}, the authors noted that geographical traffic routing could significantly reduce the use of brown energy if the price of energy was dynamically determined in proportion to the instantaneous fraction of the total energy that was brown. Moreover, the degree of use of renewable energy was dependent on the form of the pricing model and the acceptability of this dynamic energy pricing. In \cite{GaoINFOCOM13}, the authors exploited multiple uncorrelated wind energy sources to significantly reduce the effects of intermittency and to achieve almost entirely green Internet-scale data centers (IDCs). They proposed WPA, a policy that routed jobs based on the current states of the workloads and wind power availabilities at different data centers. The authors showed that more than 95\% of the energy consumption in IDCs could be satisfied by wind power without delaying the processing of jobs. Similarly, Gao \emph{et al.} \cite{GaoSIGCOMM12} dynamically controlled the fraction of user traffic directed to each geographically distributed data center with the objective of navigating the three-way trade-off among carbon footprint, access latency and electricity cost. The authors showed that carbon emissions could be reduced by 10\% without increasing either the mean latency or the electricity cost. However, at present, renewable energy is more expensive to use than is brown energy produced with fossil-based fuel. Zhang \emph{et al.} \cite{ZhangMiddleware11} considered the problem of maximizing the percentage of renewable energy used to reduce the negative environmental implications given certain operation budgets. They proposed a novel middleware system that dynamically distributed requests among different data centers. The authors formulated their objective as a constrained optimization problem and solved it via linear-fractional programming. The authors experimentally demonstrated that their system could significantly increase the use of renewable energy without violating the desired operational cost budget, despite the intermittent nature of the supplies of renewable energy at different locations and the time variation in electricity prices and workload traces.
\subsection{Future Research Directions}
To reduce the energy consumption of data centers through efficient utilization of resources, many proposals have been put forward for the green implementation data centers. Because of their exponential growth and underutilization, data centers still pose several challenges, and several inadequately addressed research directions remain with regard to resource management:
\begin{itemize}
\item A major reason why the resource utilization of data centers remains low is that operators are concerned about potential QoS violations. Recently, several studies have suggested the existence of interference between collocated applications \cite{MarsMICRO11, YangISCA13} when servers are virtualized and shared to improve utilization. The challenge of virtual machine management and data center right-sizing is to improve resource utilization while preserving QoS guarantees, which benefits both data center operators and users.
\item In a multi-tenant data center environment, different tenants may desire different levels of application performance. This characteristic requires heterogeneous resource management mechanisms, which will introduce additional complexity and overhead. Moreover, tenants possess their own resources, which may not be controlled by the data center operators in a centralized manner. Thus far, few studies have focused on energy-aware resource management in multi-tenant data centers, and the task of providing efficient polices for such scenarios requires future exploration.
\item Although traffic consolidation technologies offer considerable energy savings, they can also have severe effects on network performance. The virtualization of network functions is a promising approach to providing efficient support for the deployment and management of network services. However, current implementations such as CloudNaaS \cite{BensonAkellaSOCC11} do not account for energy consumption. Finding a good balance between energy consumption and network performance remains a challenging problem for current data center networks.
\item Most existing studies concerning power distribution are focused on intra-data-center networking. However, geographically distributed data centers can also offer opportunities for power distribution. For example, a workload can be preferentially scheduled to data centers with high power capping or a greater amount of stored energy by virtue of being located nearer to power stations. The question of how to reduce energy consumption by combining power distribution, application performance and workload scheduling remains an open research problem.
\item A demand-response mechanism is an energy-efficient method of overcoming the instability that is characteristic of renewable energy \cite{LiuSIGMETRICS11}. In SaaS data centers, such a mechanism also causes service providers to lose users in response to supply conditions. A promising solution is to design beneficial pricing schemes to incentivize service providers to reduce brown energy consumption.  
\end{itemize} 
\section{Temperature and Thermal Control}\label{sec:Temperature and thermal control}
A large portion of the energy used in computing devices is converted into heat. In chip design, the heat problem is extremely important, as it can threaten to cause the chips to fracture at extreme temperatures \cite{SkadronTACO04}. An ambient temperature range of $68^{\circ}\mathrm{F}$ to $75^{\circ}\mathrm{F}$ is optimal for device reliability \cite{GrundyAVTECH05}. Data centers house tens of thousands of servers and communication networks, which causes their cooling costs for heat removal to increase exponentially with increasing energy consumption. It is reported that more than a third and sometimes as much as one half of the total cost of a data center consists of cooling costs \cite{BeladyReport08}. As mentioned earlier, virtualizing and consolidating servers can increase processor utilization rates. To reduce management expenses and floor-space requirements, operators prefer to install many multi-core-processor server blades in a single rack-mounted enclosure. These trends toward concentration and higher-density computing in combination with the abovementioned issues mean that the problem of temperature and thermal control must play a critical role in the attempt to establish green data centers. 
\subsection{Cooling and Workload Distribution}
Optimization of the delivery of cooling is one strategy for temperature control \cite{PatelIPACK03}. Patel \emph{et al.} \cite{PatelIPACK03} proposed a system to fully control a data center environment by linking all its attributes together, including distributed sensing, variable air conditioning and data aggregation. In addition, they exploited variable opening plenum tiles to assist in achieving more flexible control of the cooling distribution and in directing cooling resources where and when needed. The authors reported an energy savings of 50\%. In \cite{ZhengHPCA14}, the authors proposed TE-Shave, a generalized framework that exploited thermal energy storage (TES) tanks to reduce data center power. TE-Shave switched to the TES for stored cold water or ice to supplement the chillers and for heat exchange during peak power periods. Moreover, TE-Shave charged the TES through the preparation of cold water or ice by increasing the cooling power when the data center experienced power valleys. The authors demonstrated that their proposed framework could lead to a 28\% savings in operational expenses compared with existing work that focused only on reducing the server-side power demand. 

Another approach to addressing temperature modulation is based on workload assignment and migration between servers to achieve thermal balance \cite{SharmaInternetComputing05, MooreATEC05, HeathASPLOS06}. In \cite{MooreATEC05}, the authors presented temperature-aware workload assignment algorithms to minimize the energy expended by the cooling infrastructure. Their algorithms produced the recommended theoretical heat distribution to maximize cooling efficiency and minimize the amount of heat recirculating within a data center. The proposed algorithms achieved cooling savings of more than 25\%. Heath \emph{et al.} \cite{HeathASPLOS06} developed Freon, a user-space system to achieve thermal management without introducing unnecessary performance degradation. The main policy driving the operation of Freon was periodic temperature monitoring and the exertion of feedback control over the utilization of the servers by distributing requests based on load weights. Generally speaking, Freon dynamically shifted load away from hot servers and increased the load on other, ``cool'' servers. An extended version named Freon-EC was proposed to combine energy conservation and thermal management. Freon-EC turned off servers based on their temperatures and physical locations in the computer room. In fact, Freon-EC turned hot servers off and replaced them with servers from different regions that were unaffected by the thermal threat. Through experiments, Freon was proven to manage temperature with as little potential throughput degradation as possible. In \cite{XuICAC13}, the authors considered the problem of temperature-aware workload distribution in geo-distributed data centers. They explored two key aspects: 1) utilizing the geographical diversity of temperature to reduce cooling energy consumption and 2) exploiting the elastic nature of batch workloads to dynamically adjust capacity allocation. The problem was formulated as a joint optimization of request routing for interactive workloads and capacity allocation for batch workloads, and it was solved using a distributed $m$-block \textbf{A}lternating \textbf{D}irection \textbf{M}ethod of \textbf{M}ultipliers (ADMM) algorithm. Extensive simulations using real-world traces revealed that their approach saved 15-20\% in cooling energy and 5-20\% in overall energy cost. In addition to these job-allocation-centric techniques, $T^*$ \cite{KaushikSC12} provided a data-centric approach to reducing cooling energy costs in big data analytics cloud computing. Files were proactively allocated in a thermal- and energy-aware manner based on knowledge of server-profile, data-semantics, and cluster information. $T^*$ proposed predictive and non-predictive temperature-aware file allocation algorithms to reduce cooling energy costs and ensure high data-local performance. Evaluations using one-month-long real-world traces demonstrated a reduction of up to 42\% in cooling energy costs and a $9\times$ better average response time compared with techniques that were data-allocation-agnostic in nature. 
\subsection{Temperature-Reliability Trade-off}
Several other efforts have focused on supplying data centers with relatively warm air to reduce cooling costs, i.e., increasing/controlling the temperature setpoint. Studies show that increasing the temperature setpoint by merely one degree can yield an energy savings of 2-5\% \cite{BrandonProcessor07}. However, raising the temperature of the server's components can reduce the system reliability. Based on the critical observation that hot spots can be selectively cooled, \cite{BiswasISCA11} proposed a cooling-power management mechanism to allow global data center coolers to operate at a higher temperature while achieving the same level of chip reliability. The authors integrated thermal models of various components, such as the silicon in the chip, embedded thermo-electric coolers, and air conditioners, and quantified the expected savings in cooling power for the entire data center. Upon evaluating over 43 applications from the SPECcup 2000 benchmark, they reported a cooling power savings of 12\% on average. However, there is a limited understanding of how higher temperatures will affect data center systems. In \cite{El-SayedSIGMETRICS12}, the authors presented a multi-faceted study of temperature management in data centers. By analyzing a large collection of field data from different production environments, the authors studied the influence of higher temperatures on 1) hardware reliability, such as storage subsystem reliability, memory subsystem reliability, and server reliability as a whole, as well as 2) server performance and power consumption. They concluded that the operators could run their data centers hotter than they currently were to save energy while still limiting the resultant negative effects on system reliability and performance.
\subsection{Open Questions and Research Directions}
The continued growth of data centers is resulting in sharply increasing levels of energy consumption required for both powering these data centers and their thermal management. Several methodologies have recently been introduced that offer promising approaches to reducing the escalating energy costs associated with meeting their cooling requirements. However, there are still several challenges and research issues that must be addressed regarding the temperature and thermal control of data centers:
\begin{itemize}
\item Most existing studies have focused on heat dissipation and cooling load management within data centers. Chillers are used to cool the hot water that returns from the Computer Room Air Conditioners (CRACs) via mechanical refrigeration cycles, and the compressors for these chillers consume large amounts of energy \cite{ZhouITherm12}. Recently, various so-called free cooling technologies, such as air-side economizing and water-side economizing, have been adopted to reduce the dependence on mechanical chillers. Air-side economizing refers to directing cold outside air into the data center to cool the servers and ejecting the hot exhaust air back outside instead of cooling and recirculating it. Water-side economizing refers to cooling down the warmer chilled water from the computer room using water streams. However, these free cooling technologies heavily depend on air conditions and geographical locations. An interesting question is whether and how these technologies can be coupled to address power and cooling demands. A related study has already been performed \cite{XuICAC13}, but more thorough theoretical and practical studies are needed to explore this possibility.
\item As mentioned earlier, service providers may own geographically distributed data centers to better meet their QoS needs, such as delay guarantees. Many of these providers, such as Google \cite{MillerGoogleThermal12}, have been relying upon fresh air for cooling to reduce the energy costs of operating their data centers. When renewable energy is considered, the problem becomes more interesting and sophisticated. For example, for systems that rely on air-side economization, it is typically preferable to operate in a cold-temperature environment. However, the generation of renewable energy using solar panels requires hot, sunny days, which will adversely affect the cooling energy consumption. Addressing this issue will require holistic methods that simultaneously consider cooling energy costs, the availability of renewable energy, and performance requirements.
\item In the current thermal control approaches implemented in most data centers, the CRAC fans are operating at their peak speed. Lowering CRAC fan speeds allows energy consumption to be reduced. However, the challenge is to develop CRAC fan speed controllers that can supply a data center with only the necessary cooling demands. The existing work of \cite{SundaralingamASME11} represents an initial effort to address the problem of fan speed management.
\end{itemize}
\section{Green Metrics, Monitoring and Experimental Techniques}\label{sec:Green metrics, monitoring and experimental techniques}
In this section, we present the metrics relevant to building a ``green'' data center and practical efforts to monitor and test data centers for the optimization of energy consumption and carbon emissions. The primary goals of developing green metrics are to 1) quantitatively describe the degree of greenness of a data center, 2) track and report the energy usage and carbon emissions of a data center to increase its green efficiency, and 3) provide a clear visualization of the efficiency of a data center to help its operators to reduce their electricity bills and carbon taxes. In the meantime, monitoring efforts provide data center operators with a real-time, detailed understanding of operation parameters, including green metrics. As a result, operators can adopt appropriate green measures based on the output parameters, for example, the consolidation of virtual machines when the monitoring data indicate that the data center is experiencing low utilization. Moreover, experimentation and simulation techniques are also critical for designing test tools and demonstrating energy-efficient policies.
\subsection{Green Data Center Metrics}

\begin{figure*}[htbp]
\centering
\tikzset{
box/.style={rectangle, rounded corners, draw=black, align=center, inner sep=2pt, minimum height=1cm}
}
\begin{tikzpicture}[node distance=2cm] 
 \node[box, line width=0.9pt] (n1) at (0,2) {Set benchmarking goals};
 \node[box, line width=0.9pt] (n2) at (4,2) {Prioritize metrics};
 \node[box, line width=0.9pt] (n3) at (8.6,2) [text width=5cm] {Identify required data and develop data collection plan};
 \node[box, line width=0.9pt] (n4) at (13.7,2) [text width=4cm] {Obtain and install monitoring equipment};
 \node[box, line width=0.9pt] (n5) at (14.2,0) {Collect data};
 \node[box, line width=0.9pt] (n6) at (11,0) [text width=3cm] {Analyze data and compute metrics};
 \node[box, line width=0.9pt] (n7) at (6.9,0) [text width=4cm] {Benchmark metrics and identify potential actions};
 \node[box, line width=0.9pt] (n8) at (2.9,0) [text width=2.5cm] {Create follow-up action plan};
 \node[box, line width=0.9pt] (n9) at (0,0) {Share results};
 
 \draw [->, >=latex,line width=0.9pt] (n1) -- (n2);
 \draw [->, >=latex,line width=0.9pt] (n2) -- (n3);
 \draw [->, >=latex,line width=0.9pt] (n3) -- (n4);
 \draw [->, >=latex,line width=0.9pt] (14.2,1.5) -- (14.2,0.5);
 \draw [->, >=latex,line width=0.9pt] (n5) -- (n6);
 \draw [->, >=latex,line width=0.9pt] (n6) -- (n7);
 \draw [->, >=latex,line width=0.9pt] (n7) -- (n8);
 \draw [->, >=latex,line width=0.9pt] (n8) -- (n9);
\end{tikzpicture}
\caption{\label{fig:Benchmarking_metrics_process} The benchmarking process for data center metrics \cite{MathewLBNL09}}
\end{figure*}
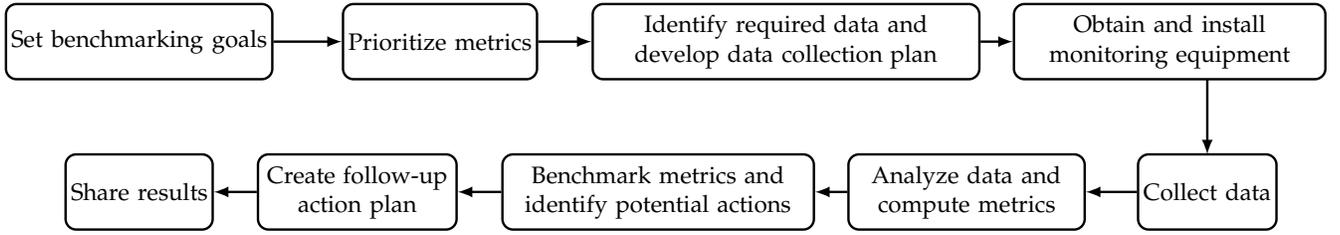
Various metrics focused on data center energy efficiency have arisen over the past years. We summarize the most popular green metrics below:
\subsubsection{Power Usage Effectiveness (PUE)}
\begin{eqnarray*}
PUE=\frac{Total\;Facility\;Power}{IT\;Equipment\;Power}
\end{eqnarray*}
The IT equipment power is the power consumed by the servers, network, storage, and supplemental equipment (i.e., workstations/laptops used for monitoring); the total facility power includes the aforementioned power plus auxiliary consumption in the data center \cite{BeladyGreenGrid08}.
\subsubsection{Data Center infrastructure Efficiency (DCiE)}
\begin{eqnarray*}
DCiE=\frac{1}{PUE}=\frac{IT\;Equipment\;Power}{Total\;Facility\;Power}
\end{eqnarray*}
This metric is simply the inverse of the PUE \cite{BeladyGreenGrid08}, and both were developed by The Green Grid to measure energy efficiency in data centers. Barroso \emph{et al.} \cite{Barroso13} suggested that the average value of the PUE is 1.83, i.e., the DCiE is 0.54.
\subsubsection{Data Center energy Productivity (DCeP)}
\begin{eqnarray*}
DCeP=\frac{Useful\;Work\;Produced}{Total\;Energy\;Consumed\;to\;Perform\;that\;Work}
\end{eqnarray*}
This is a metric to characterize the computing efficiency of a data center \cite{AndersonGreenGrid08}. Useful work refers to the jobs performed by the IT equipment during the time period. The denominator is the total energy consumed during the assessment window.
\subsubsection{Heating, Ventilation, and Air Conditioning (HVAC)}
\begin{eqnarray*}
HVAC=\frac{IT\;Electrical\;Energy}{HVAC+(Fuel+Steam+Chilled){\times}293}
\end{eqnarray*}
The HVAC system energy is the sum of the electrical energy used for cooling, fan movement, and any other HVAC energy usage such as steam or chilled water, and it can be reduced via free cooling with air economizers and/or high-efficiency chillers \cite{MathewLBNL09}. When the value of this metric is high, it suggests a high potential to reduce HVAC energy usage.
\subsubsection{Carbon Usage Effectiveness (CUE)}
\begin{eqnarray*}
CUE=\frac{Total\;CO_2\;Emissions}{IT\;Equipment\;Energy}
\end{eqnarray*}
The total carbon emissions include direct greenhouse gas emissions and indirect greenhouse gas emissions from the consumption of purchased electricity, heat, and steam. The denominator is the energy consumption of the IT equipment \cite{AzevedoGreenGrid10}.

In addition to these metrics, many other metrics have been proposed by The Green Grid and Lawrence Berkeley National Laboratory that have different practical uses, such as the CEF (Carbon Emission Factor), AEU (Air Economizer Utilization), WEU (Water Economizer Utilization) and CSE (Cooling System Efficiency) \cite{MathewLBNL09, AzevedoGreenGrid10}. Table~\ref{tb:Green metrics taxonomy} presents the taxonomy of the green metrics listed above. Note that the present values shown in the table were obtained from \cite{MathewLBNL09, AzevedoGreenGrid10} if no other reference is indicated.
\renewcommand{\arraystretch}{1.3}
\begin{table*}[htbp]
\centering
\scriptsize
\caption{\label{tb:Green metrics taxonomy} Green metrics taxonomy}
\begin{tabular}{p{4.6cm}|p{4.4cm}|p{2.5cm}|p{1.0cm}|p{2.3cm}}
\hline\hline
Metric Definition & Objective & Proposer & Ideal & Present Value  \\\hline
\multicolumn{1}{c|}{\multirow{2}{4.6cm}{$PUE=\frac{Total\;Facility\;Power}{IT\;Equipment\;Power}$}} & 
\multicolumn{1}{c|}{\multirow{2}{4.4cm}{Characterize the total energy efficiency of a data center}} & 
\multicolumn{1}{c|}{\multirow{2}{2.8cm}{Green Grid, 2008}} & 
\multicolumn{1}{c|}{\multirow{2}{1.0cm}{{$\rightarrow$}1}} & 
\multicolumn{1}{c}{\multirow{2}{2.3cm}{1.83}} \\
\multicolumn{1}{c|}{}   &   \multicolumn{1}{c|}{}   &   \multicolumn{1}{c|}{}   & \multicolumn{1}{c|}{}  & \multicolumn{1}{c}{} \\

\hline
\multicolumn{1}{c|}{\multirow{2}{4.6cm}{$DCiE=\frac{1}{PUE}=\frac{IT\;Equipment\;Power}{Total\;Facility\;Power}$}} & 
\multicolumn{1}{c|}{\multirow{2}{4.4cm}{Characterize the total energy efficiency of a data center}} & 
\multicolumn{1}{c|}{\multirow{2}{2.8cm}{Green Grid, 2008}} & 
\multicolumn{1}{c|}{\multirow{2}{1.0cm}{{$\rightarrow$}1}} &
\multicolumn{1}{c}{\multirow{2}{2.3cm}{0.54}} \\
\multicolumn{1}{c|}{}   &   \multicolumn{1}{c|}{}   &   \multicolumn{1}{c|}{}   & \multicolumn{1}{c|}{}  & \multicolumn{1}{c}{} \\

\hline
\multicolumn{1}{c|}{\multirow{2}{4.6cm}{$DCeP=\frac{Useful\;Work\;Produced}{Total\;Energy\;Consumed\;to\;Perform\;that\;Work}$}} & 
\multicolumn{1}{c|}{\multirow{2}{4.4cm}{Characterize the IT computing efficiency}} & 
\multicolumn{1}{c|}{\multirow{2}{2.8cm}{Green Grid, 2008}} & 
\multicolumn{1}{c|}{\multirow{2}{1.0cm}{Larger}} &
\multicolumn{1}{c}{\multirow{2}{2.3cm}{866658$(\frac{Normalized Tasks}{KWhr})$}} \\
\multicolumn{1}{c|}{}   &   \multicolumn{1}{c|}{}   &   \multicolumn{1}{c|}{}   & \multicolumn{1}{c|}{}  & \multicolumn{1}{c}{} \\

\hline
\multicolumn{1}{c|}{\multirow{2}{4.6cm}{$HVAC=\frac{IT\;Electrical\;Energy}{HVAC+(Fuel+Steam+Chilled){\times}293}$}} & 
\multicolumn{1}{c|}{\multirow{2}{4.4cm}{Characterize the energy efficiency of the HVAC system}} & 
\multicolumn{1}{c|}{\multirow{2}{2.9cm}{Lawrence Berkeley National Laboratory, 2009}} & 
\multicolumn{1}{c|}{\multirow{2}{1.0cm}{Higher}} &
\multicolumn{1}{c}{\multirow{2}{2.3cm}{-\,-}} \\
\multicolumn{1}{c|}{}   &   \multicolumn{1}{c|}{}   &   \multicolumn{1}{c|}{}   & \multicolumn{1}{c|}{}  & \multicolumn{1}{c}{} \\

\hline
\multicolumn{1}{c|}{\multirow{2}{4.6cm}{$CEF=\frac{Total\;CO_2\;Emissions}{Total\;Facility\;Energy}$}} & 
\multicolumn{1}{c|}{\multirow{2}{4.4cm}{Assess the carbon emissions per unit of energy used}} & 
\multicolumn{1}{c|}{\multirow{2}{2.9cm}{Green Grid, 2010}} & 
\multicolumn{1}{c|}{\multirow{2}{1.0cm}{Lower}} &
\multicolumn{1}{c}{\multirow{2}{2.3cm}{0.147 (Washington) \cite{USAEIAReport07}}} \\
\multicolumn{1}{c|}{}   &   \multicolumn{1}{c|}{}   &   \multicolumn{1}{c|}{}   & \multicolumn{1}{c|}{}  & \multicolumn{1}{c}{} \\

\hline
\multicolumn{1}{c|}{\multirow{2}{4.6cm}{$CUE=CEF{\times}PUE=\frac{Total\;CO_2\;Emissions}{IT\;Equipment\;Energy}$}} & 
\multicolumn{1}{c|}{\multirow{2}{4.4cm}{Characterize the overall efficiency of the cooling system}} & 
\multicolumn{1}{c|}{\multirow{2}{2.9cm}{Green Grid, 2010}} & 
\multicolumn{1}{c|}{\multirow{2}{1.0cm}{Lower}} &
\multicolumn{1}{c}{\multirow{2}{2.3cm}{0.27 (Washington)}} \\
\multicolumn{1}{c|}{}   &   \multicolumn{1}{c|}{}   &   \multicolumn{1}{c|}{}   & \multicolumn{1}{c|}{}  & \multicolumn{1}{c}{} \\

\hline
\multicolumn{1}{c|}{\multirow{2}{4.6cm}{$CSE=\frac{Average\;Cooling\;System\;Power\;Usage}{Average\;Cooling\;Load}$}} & 
\multicolumn{1}{c|}{\multirow{2}{4.4cm}{Represent the carbon emission efficiency of IT energy use}} & 
\multicolumn{1}{c|}{\multirow{2}{2.9cm}{Lawrence Berkeley National Laboratory, 2009}} & 
\multicolumn{1}{c|}{\multirow{2}{1.0cm}{Lower}} &
\multicolumn{1}{c}{\multirow{2}{2.3cm}{0.6{-}0.8$(\frac{KW}{Ton})$}} \\
\multicolumn{1}{c|}{}   &   \multicolumn{1}{c|}{}   &   \multicolumn{1}{c|}{}   & \multicolumn{1}{c|}{}  & \multicolumn{1}{c}{} \\

\hline
\multicolumn{1}{c|}{\multirow{3}{4.6cm}{$AEU=\frac{Air\;Economizer\;Hours}{24{\times}365}$}} & 
\multicolumn{1}{c|}{\multirow{3}{4.4cm}{Measure the percentage of hours in a year that an air{-}side economizer system is used to provide ``free cooling''}} & 
\multicolumn{1}{c|}{\multirow{3}{2.9cm}{Lawrence Berkeley National Laboratory, 2009}} & 
\multicolumn{1}{c|}{\multirow{3}{1.0cm}{{$\rightarrow$}$100\%$}} &
\multicolumn{1}{c}{\multirow{3}{2.3cm}{$76\%$ (New York)}} \\
\multicolumn{1}{c|}{}   &   \multicolumn{1}{c|}{}   &   \multicolumn{1}{c|}{}   & \multicolumn{1}{c|}{}  & \multicolumn{1}{c}{} \\
\multicolumn{1}{c|}{}   &   \multicolumn{1}{c|}{}   &   \multicolumn{1}{c|}{}   & \multicolumn{1}{c|}{}  & \multicolumn{1}{c}{} \\

\hline
\multicolumn{1}{c|}{\multirow{3}{4.6cm}{$WEU=\frac{Water\;Economizer\;Hours}{24{\times}365}$}} & 
\multicolumn{1}{c|}{\multirow{3}{4.4cm}{Measure the percentage of hours in a year that a water{-}side economizer system is used to provide ``free cooling''}} & 
\multicolumn{1}{c|}{\multirow{3}{2.9cm}{Lawrence Berkeley National Laboratory, 2009}} & 
\multicolumn{1}{c|}{\multirow{3}{1.0cm}{{$\rightarrow$}$100\%$}} &
\multicolumn{1}{c}{\multirow{3}{2.3cm}{$33\%=\frac{2900}{24{\times}365}$ \cite{42UWEU11}}} \\
\multicolumn{1}{c|}{}   &   \multicolumn{1}{c|}{}   &   \multicolumn{1}{c|}{}   & \multicolumn{1}{c|}{}  & \multicolumn{1}{c}{} \\
\multicolumn{1}{c|}{}   &   \multicolumn{1}{c|}{}   &   \multicolumn{1}{c|}{}   & \multicolumn{1}{c|}{}  & \multicolumn{1}{c}{} \\
\hline\hline
\end{tabular}
\end{table*}

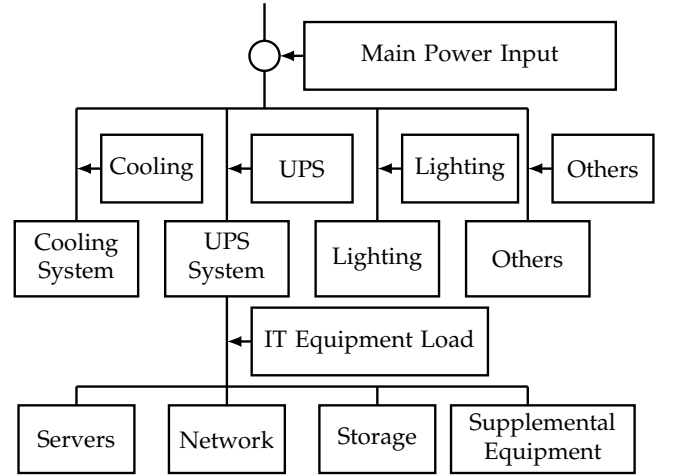
\begin{figure}[htbp]
\centering
\tikzset{
box/.style={rectangle, draw=black, align=center, inner sep=2pt}
}
\begin{tikzpicture}
  \draw[line width=0.9pt] (2.5,6.4)--(2.5,5.9);
  \draw[line width=0.9pt] (2.5,5.7) circle (0.2);
  \draw[line width=0.9pt] (2.5,5.5)--(2.5,5);
  \draw[line width=0.9pt] (0,5)--(6,5);
  
  \node[box, line width=0.9pt] (n10) at (5.1,5.7) [text width=4cm, minimum height=0.9cm]{Main Power Input};
  
  \draw [->, >=latex,line width=0.9pt] (3.02,5.7) -- (2.7,5.7);
  
  \draw[line width=0.9pt] (0,5)--(0,3.5);
  \draw[line width=0.9pt] (2,5)--(2,3.5);
  \draw[line width=0.9pt] (4,5)--(4,3.5);
  \draw[line width=0.9pt] (6,5)--(6,3.5);
  
  \node[box, line width=0.9pt] (n1) at (0,3) [text width=1.5cm, minimum height=1cm]{Cooling System};
  \node[box, line width=0.9pt] (n2) at (2,3) [text width=1.5cm, minimum height=1cm]{UPS System};
  \node[box, line width=0.9pt] (n3) at (4,3) [text width=1.5cm, minimum height=1cm]{Lighting};
  \node[box, line width=0.9pt] (n4) at (6,3) [text width=1.5cm, minimum height=1cm]{Others};
  
  \node[box, line width=0.9pt] (n5) at (1,4.2) [text width=1.2cm, minimum height=0.9cm]{Cooling};
  \node[box, line width=0.9pt] (n6) at (3,4.2) [text width=1.2cm, minimum height=0.9cm]{UPS};
  \node[box, line width=0.9pt] (n7) at (5.1,4.2) [text width=1.4cm, minimum height=0.9cm]{Lighting};
  \node[box, line width=0.9pt] (n8) at (7,4.2) [text width=1.2cm, minimum height=0.9cm]{Others};
  
  \draw [->, >=latex,line width=0.9pt] (0.32,4.2) -- (0,4.2);
  \draw [->, >=latex,line width=0.9pt] (2.32,4.2) -- (2,4.2);
  \draw [->, >=latex,line width=0.9pt] (4.32,4.2) -- (4,4.2);
  \draw [->, >=latex,line width=0.9pt] (6.32,4.2) -- (6,4.2);
  
  \draw[line width=0.9pt] (2,2.5)--(2,1.3);
  \node[box, line width=0.9pt] (n9) at (3.9,1.9) [text width=3cm, minimum height=0.9cm]{IT Equipment Load};
  
  \draw [->, >=latex,line width=0.9pt] (2.32,1.9) -- (2,1.9);
  
  \draw[line width=0.9pt] (0,1.3)--(6,1.3);
  
  \node[box, line width=0.9pt] (n11) at (0,0.6) [text width=1.4cm, minimum height=0.9cm]{Servers};
  \node[box, line width=0.9pt] (n12) at (2,0.6) [text width=1.4cm, minimum height=0.9cm]{Network};
  \node[box, line width=0.9pt] (n13) at (4,0.6) [text width=1.4cm, minimum height=0.9cm]{Storage};
  \node[box, line width=0.9pt] (n14) at (6.2,0.6) [text width=2.3cm, minimum height=0.9cm]{Supplemental Equipment};
  
  \draw[line width=0.9pt] (0,1.3)--(0,1.05);
  \draw[line width=0.9pt] (2,1.3)--(2,1.05);
  \draw[line width=0.9pt] (4,1.3)--(4,1.05);
  \draw[line width=0.9pt] (6,1.3)--(6,1.05);
\end{tikzpicture}
\caption{\label{fig:Power_measuring_point_registers} Power measuring point registries}
\end{figure}

Now, we focus on works that have studied how to measure these metrics. A technical report \cite{MathewLBNL09} provided an approach for obtaining the particular values of the metrics. The specific implementation process for benchmarking a data center is illustrated in Fig.~\ref{fig:Benchmarking_metrics_process}. The authors of this report also provided guidance for calculating the required data. For example, electrical energy use can be derived from meter data or utility bills. In \cite{JaureguialzoINTELEC11}, the authors' intent was to present the ``what'' and ``how'' for the measurement of metrics of energy efficiency. They attempted to introduce a cheap and effective continuous measurement method based on the power demand profiles recorded at various points in a facility. Their power measuring point system is illustrated in Fig.~\ref{fig:Power_measuring_point_registers}. When the required data have been obtained from the various registries, the metrics can be calculated based on the power consumed integrated over 15-minute or 1-hour periods. By using standard measuring equipment, the reliability and accuracy of the results can be guaranteed. Another work \cite{GaoReportGoogle14} adopted machine learning techniques to study the PUE. The authors developed a neural network framework that learned from actual operations data to predict the PUE. They reported that the PUE was within 0.004+/-0.005, with an error of approximately $0.4\%$ for a PUE of 1.1. DCeP, the metric of energy productivity, was studied in \cite{SegoJETC12}. The authors presented an approach to calculating the DCeP by designing an experiment using a highly instrumented, high-performance computing data center. Their evaluation results demonstrated that the DCeP could be used to successfully distinguish among different operational states in the data center and validated its utility as a metric for identifying the configurations of hardware and software that could improve energy productivity. Unfortunately, the improvement of one specific green metric may lead to the worsening of another metric. Therefore, holistic frameworks for addressing all metrics have also been proposed in recent works \cite{GIPCReport12, USAEuropeanJapan14}. These approaches offer a combined visualization of all metrics to guide data center operators in worst-metric improvement.

\subsection{Green Monitoring and Experimental Technique}
Monitoring is an effective method of detecting and proactively mitigating the failure of data centers. Moreover, we will explore efforts that yield the tracks and traces of energy usage, thermal emission, and power distribution for individual devices. Based on the collected information, intelligent mechanisms can then react to increase the energy efficiency and productivity of a data center. In \cite{LiuICACINDST09}, the authors presented a comprehensive online monitoring service in their GreenCloud architecture. The monitoring service monitored and collected data concerning resource utilization, application workloads directed through virtual machine hypervisors, and power usage using an intelligent power distribution unit. Analyzing and mining this collected information provided a clear understanding of the data center energy usage and temperature behavior. Furthermore, the monitoring service helped to consolidate the virtual machine workload and achieve significant energy savings while guaranteeing the real-time performance required for sensitive applications. Using a different approach, the authors of \cite{LuICAC12} developed PowerTracer, a request-tracing strategy for diagnosing energy inefficiencies in multi-tier service systems. They first profiled the patterns of requests through kernel instrumentation and then measured the server-side latency and service time of each tier. Finally, PowerTracer collected the resource consumption information for individual requests and analyzed the features of each tier and each pattern. By thus gaining insights into the root causes of energy inefficiency, the authors could devise efficient power-saving methods for multi-tier applications.

Even in the initial phases of the design and development of a data center, an effective experimental and simulation platform enables the operator to identify and understand the design aspects that are critical to energy and resource efficiency. EEFSim, a virtualized data center simulator for cloud computing research, was designed and developed in \cite{JuliaTechniqueReport10}. The simulator was demonstrated to be suitable for testing, as it reproduced the behavior of a real cloud framework and the execution information that it offered. Another advantage of this simulator was that it could be easily used to evaluate the power performance of a system by testing different solutions and approaches within merely a few minutes. As a result, the simulator offered the possibility to validate different scheduling and migration policies for their power consumption associated with virtual machines in a cloud data center. Eliazovich \emph{et al.} \cite{KliazovichGlobecom10} presented GreenCloud, a packet-level simulator for energy-aware studies of cloud computing data centers. This simulator could be used to capture the details of the energy consumption of various data center components (e.g., servers, switches, and links) as well as packet-level communication patterns. Its effectiveness in utilizing power management schemes was demonstrated for two-tier, three-tier, and three-tier high-speed data center architectures. A more powerful tool for analyzing green data center design and resource management, named GDCSim, was proposed in \cite{GuptaIGCC11}. This simulation tool was used for the iterative design of green data centers. Specifically, it captured the interdependencies between online resource management schemes and the physical behavior of data centers. The objective of the simulator was to generate data center configurations for particular purposes, such as CPU sleep-state transitions and DVFS for power management, as well as thermal behavior characterization. In \cite{YeoComputer12}, the authors proposed a simulator, named SimWare, to analyze the power consumption of servers, cooling units, and fans as well as the effects of heat recirculation and air supply timing. SimWare considered the time required for air to travel from the CRACs to the servers and could evaluate user-defined job scheduling and virtual machine management algorithms. Data center designers could use this simulator to evaluate energy-conserving policies and to assess mechanical design options such as server placement, airflow management and CRAC control strategies.
\subsection{Challenges and Design Guidelines}
Currently, the PUE is a widely accepted and measured metric in data centers, whereas the use of other green metrics is much less widespread. There is no doubt that green efficiency metrics will continue to be developed and drive improvements in sustainability and environmental friendliness. However, an effective metric to characterize the energy and resource usage efficiency of a data center must satisfy the following design guidelines:
\begin{itemize}
\item able to track the important and actionable parameters to assist operators in making data centers more green;
\item consistent, easy and cost effective to implement and report for management purposes;
\item feasible, simple and inexpensive to monitor.
\end{itemize}
Thus, developing metrics that are well suited for reducing energy consumption is a challenging task with a significant contribution to the construction of ``green'' data centers.

Moreover, the monitoring of green metrics in data centers assists in diagnosing energy inefficiencies and in the implementation of proactive mechanisms to improve energy efficiency. However, this issue has received little attention thus far, and existing monitors have been proposed for either particular architectures \cite{LiuICACINDST09} or special applications \cite{LuICAC12}. The challenges are the large number of hosted servers or virtual machines and the availability of communication resources in data centers. Consequently, centralized monitoring techniques suffer from low scalability and elasticity. Monitoring techniques adopted from cooperative monitoring \cite{XuConfIPCCC08} can be applied to overcome these drawbacks by enabling distributed and robust monitoring solutions for large-scale, complex data centers. The key research question of interest is how to minimize the costs incurred while guaranteeing monitoring accuracy.

Finally, although platforms such as GreenCloud \cite{KliazovichGlobecom10} and GDCSim \cite{GuptaIGCC11} have provided data center energy-aware simulation environments, the primary disadvantage of these platforms is that most of them are coarse grained and focused only on certain components or functions. For example, when a simulation of virtual machine migration considers only CPU resources, the results for virtual machine allocation may suffer from a failure to consider the bandwidth requirements. Therefore, the task of designing a system-wide experimental and simulation platform that integrates all components of a data center, such as the CPU, memory, cache, I/O, disk and communication network, requires further exploration.  
\section{Conclusion}\label{sec:Conclusion}
The development of green data centers has garnered considerable attention because of their financial and environmental impact. Extensive research is being conducted in an attempt to reduce energy consumption and carbon emissions. In this article, we present a comprehensive review of the current state-of-the-art research in the area of green data center management. This survey offers significant insights into various aspects of the problem, including energy efficiency, resource management, thermal control and green metrics. We summarize and compare the existing schemes and highlight the challenges and potential directions for future research with regard to each of these aspects.

\end{document}